\documentclass[pre,aps,12pt,superscriptaddress]{revtex4}

\usepackage{graphicx}
\usepackage{appendix}
\usepackage{amsmath,amssymb}

\newcommand{\degree}{\ensuremath{^\circ}}
\newcommand{\ignorar}[1]{}

\begin{document}

\title{The reduction of plankton biomass induced by mesoscale
stirring: a modeling study in the Benguela upwelling.}

\author{Ismael Hern\'andez-Carrasco} 
\author{Vincent Rossi} 
\author{Crist\'obal L\'opez}
\author{Emilio Hern\'andez-Garc\'ia} 
\affiliation{
IFISC, Instituto de F\'isica Interdisciplinar y Sistemas Complejos (CSIC-UIB),
07122 Palma de Mallorca, Spain}

\author{Veronique Gar\c{c}on}
\affiliation{Laboratoire d'\'Etudes en G\'eophysique et 
Oc\'eanographie Spatiale, CNRS, Observatoire Midi-Pyr\'en\'ees, 14 avenue 
Edouard Belin, Toulouse, 31401 Cedex 9, France}

\date{\today}

\begin{abstract}

Recent studies, both based on remote sensed data and coupled
models, showed a reduction of biological productivity due to
vigorous horizontal stirring in upwelling areas. In order to
better understand this phenomenon, we consider a system
of oceanic flow from the Benguela area coupled with a simple
biogeochemical model of Nutrient-Phyto-Zooplankton (NPZ) type.
For the flow three different surface velocity fields are
considered: one derived from satellite altimetry data, and the
other two from a regional numerical model at two different
spatial resolutions. We compute horizontal particle dispersion
in terms of Lyapunov Exponents, and analyzed their correlations
with phytoplankton concentrations. Our modelling approach
confirms that in the south Benguela there is a reduction of
biological activity when stirring is increased. Two-dimensional
offshore advection and latitudinal difference in Primary 
Production, also mediated by the flow, seem to be the 
dominant processes involved. We estimate that mesoscale 
processes are responsible for 30 to 50\% of the offshore fluxes 
of biological tracers. In the northern area, other factors
not taken into account in our simulation are influencing the 
ecosystem. We suggest explanations for these results in the 
context of studies performed in other eastern boundary upwelling 
areas.

\end{abstract}

\maketitle

\section{Introduction}
\label{Sec:Intro} Marine ecosystems of the Eastern Boundary
Upwelling zones are well known for their major contribution to
the world ocean productivity. They are characterized by
wind-driven upwelling of cold nutrient-rich waters along the 
coast that supports elevated plankton and pelagic fish
production \citep{Mackas2006}. Variability is introduced by
strong advection along the shore, physical forcings by local
and large scales winds, and high submeso- and meso-scale activities
over the continental shelf and beyond, linking the coastal
domain with the open ocean.

The Benguela Upwelling System (BUS) is one of the four major
Eastern Boundary Upwelling Systems (EBUS) of the world. The
coastal area of the Benguela ecosystem extends from southern
Angola (around 17$\degree$S) along the west coast of Namibia
and South Africa (36$\degree$S). It is surrounded by two 
boundary currents, the warm Angola Current in the north,
and the temperate Agulhas Current in the south. The BUS can itself be
subdivided into two subdomains by the powerful Luderitz
upwelling cell \citep{Hutchings2009}. Most of the
biogeochemical activity occurs within the upwelling front and
the coast, although it can be extended further offshore toward
the open ocean by the numerous filamental structures developing
offshore \citep{Monteiro2008}. In the BUS, as in the other
major upwelling areas, high mesoscale activity due to eddies
and filaments impacts strongly marine planktonic ecosystem over
the continental shelf and beyond
\citep{Brink1991,Martin2003,Sandulescu2008,Rossi2009}.

The purpose of this study is to analyze the impact of 
horizontal stirring on phytoplankton dynamics in the BUS
within an idealized two dimensional modelling framework. Based
on satellite data of the ocean surface,
\citet{Rossi2008,Rossi2009} recently suggested that mesoscale
activity has a negative effect on chlorophyll standing stocks
in the four EBUS. This was obtained by correlating remote sensed
chlorophyll data with a Lagrangian measurement of lateral
stirring in the surface ocean (see Methods section). This
result was unexpected since mesoscale physical structures,
particularly mesoscale eddies, have been related to higher
planktonic production and stocks in the open ocean
\citep{mcgillicuddy2007} as well as off a major EBUS
\citep{Correaramirez2007}. A more recent and thorough study
performed by \cite{Gruber2011} in the California and the Canary
current systems extended the initial results from
\citet{Rossi2008,Rossi2009}. Based on satellite derived
estimates of net Primary Production, of upwelling strength and
of Eddy Kinetic Energy (EKE) as a measure the intensity of
mesoscale activity, they confirmed the suppressive effect of
mesoscale structures on biological production in upwelling
areas. Investigating the mechanism behind this observation
by means of on 3D eddy-resolving coupled models, \cite{Gruber2011}
showed that mesoscale eddies tend to export offshore and
downward a certain pool of nutrients not being effectively used
by the biology in the coastal areas. This process they called
"nutrients leakage" is also having a negative feedback 
by diminishing the pool of deep nutrients available in the
surface waters being re-upwelled continuously.

In our work, we focused on the Benguela area, being the most
contrasting area of all EBUS in terms of stirring intensity
\citep{Rossi2009}. Although the mechanisms studied by
\cite{Gruber2011} seem to involve 3D dynamics, the initial
observation of this suppressive effect was essentially based on
two-dimensional (2D) datasets \citep{Rossi2008}. In this work
we use 2D numerical analysis in a semi-realistic framework to
better understand the effects of a 2D turbulent flow on
biological dynamics, apart from the complex 3D bio-physical
processes. The choice of this simple horizontal numerical
approach is indeed supported by other theoretical 2D studies
that also displayed a negative correlation between stirring and
biomass \citep{Tel2005,MacKiver2009,Neufeld2009}. Meanwhile,
since biological productivity in upwelling areas rely on the
(wind-driven) vertical uplift of nutrients, we introduced in
our model a nutrient source term with an intensity and spatial
distribution corresponding to the upwelling characteristics.
Instead of the commonly used EKE, which is an Eulerian diagnostic tool, we
used here a Lagrangian measurement of mesoscale stirring that
has been demonstrated as a powerful tool to study patchy
chlorophyll distributions influenced by dynamical structures at
mesoscale, such as upwelling filaments \citep{Calil2010}. 
The Lagrangian perspective provides a complementary insight to transport 
phenomena in the ocean with respect to the Eulerian one. In particular, 
the concept of Lagrangian Coherent Structure may give a global idea of 
transport in a given area, separating regions with different dynamical 
behavior, and signaling avenues and barriers to transport, which are of 
great relevance for the marine biological dynamics. While the Eulerian 
approach describes the characteristics of the velocity field, the 
Lagrangian one addresses the effects of this field on transported 
substances, which is clearly more directly related to the biological 
dynamics. For example the work by \cite{HernandezCarrasco2011b} describes currents 
in the world ocean having the same level of Eddy Kinetic Energy but 
having two different stirring characteristics, as quantified by 
Lagrangian tools. Further discussions comparing Lagrangian and Eulerian 
diagnostics can be found, for example, in \cite{dOvidio2009} and the 
above cited \cite{HernandezCarrasco2011b}.
To consider velocity fields with different characteristics and
to test the effect of the spatial resolution, different flow
fields are used, one derived from satellite and two produced by
numerical simulations at two different spatial resolutions. Our
modelled chlorophyll-a concentrations are compared with observed
distributions of chlorophyll-a (a metric for phytoplankton)
obtained from the SeaWiFS satellite sensor.

This paper is organized as follows. Sec. \ref{sec:data} is a
brief description of the different datasets used in this study.
Sec. \ref{sec:metodo} depicts the methodology, which includes
the computation of the finite-size Lyapunov exponents, and the
numerical plankton-flow 2D coupled model. Then, our results
are analyzed and discussed in Sec. \ref{sec:results}. Finally 
in Sec. \ref{sec:summary}, we summed-up our main findings.


\section{Satellite and simulated data}
\label{sec:data}

We used three different 2D surface velocity fields of the Benguela area. 
Two are obtained from the numerical model Regional Ocean Model System (ROMS),
and the other one from a combined satellite product.

\subsection{Surface velocity fields derived from regional simulations.}

ROMS is a free surface, hydrostatic, primitive equation model, and we used here
an eddy-resolving climatologically forced run provided by \citep{Gutknecht2013}. 
At each grid point, linear horizontal resolution is the same in both 
the longitudinal, $\phi$, and latitudinal, $\theta$, directions, 
which leads to angular resolutions $\Delta \phi =\Delta_0$ and 
$\Delta \theta = \Delta \phi \cos{\theta}$. The numerical model was run onto
2 different grids: a coarse one at spatial resolution of $\Delta_0 =1/4 \degree$, 
and a finer one at $\Delta_0 =1/12 \degree$ of spatial resolution.
In the following we label the dataset from the coarser
resolution run as \textit{ROMS1/4}, and the finer one as
\textit{ROMS1/12}. For both runs, vertical resolution is
variable with $30$ layers in total, while only data from the
surface upper layer are used in the following. Since the 
flows are obtained from climatological forcings, they would 
represent a mean annual cycle of the typical surface currents 
of the Benguela region.

\subsection{Surface velocity field derived from satellite}
A velocity field derived from satellite observations is compared 
to the simulated fields described previously. It consists of 
surface currents computed from a combination of wind-driven 
Ekman currents, at 15 m depth, derived from Quickscat 
wind estimates, and geostrophic currents calculated 
using time-variable Sea Surface Heights (SSH) obtained 
from satellite \citep{Sudre2008}. These SSH were calculated 
from mapped altimetric sea level anomalies
combined with a mean dynamic topography. This velocity field,
labeled as \textit{Satellite1/4}, covers a period from June
2002 to June 2005 with a spatial resolution of $\Delta_0 =1/4
\degree$ in both longitudinal and latitudinal directions.

\subsection{Ocean color as a proxy for phytoplankton biomass}
To validate simulated plankton concentrations, we use a
three-year-long time series, from January 2002 to January 2005,
of ocean color data. Phytoplankton pigment concentration
(chlorophyll-a) is obtained from monthly Sea viewing Wide 
Field-of-view Sensor (SeaWiFS) products, generated by the NASA
Goddard Earth Science (GES)/Distributed Active Archive Center
(DAAC). Gridded global data were used with a resolution of
approximately 9 by 9 km.

\section{Methodology}
\label{sec:metodo}

\subsection{Finite-Size Lyapunov Exponents (FSLEs)}
\label{fsle}

FSLEs \citep{Artale1997,Aurell1997,Boffetta2001} provides a
measure of dispersion, and thus of stirring and mixing, as a
function of the spatial resolution. This Lagrangian tool allows
isolating the different regimes corresponding to different
length scales of the oceanic flows, as well as identifying
Lagrangian Coherent Structures (LCSs) present in the data
\citep{TewKai2009}. FSLE are computed from $\tau$, the time
required for two particles of fluid (one of them placed at
$\textbf{x}$) to separate from an initial distance of $\delta_0$ 
(at time $t$) to a final distance of $\delta_f$, as
\begin{equation}
\lambda (\textbf{x},t, \delta_0, \delta_f)= \frac{1}{\tau}
\log{\frac{\delta_f}{\delta_0}}.
\label{formFSLE}
\end{equation}
It is natural to choose the initial points $\textbf{x}$ on the
nodes of a grid with lattice spacing coinciding with the
initial separation of fluid particles $\delta_0$. Then, values
of $\lambda$ are obtained in a grid with lattice separation
$\delta_0$. In most of this work 
the resolution of the FSLE field, $\delta_0$, is chosen equal to
the resolution of the velocity field, $\Delta_0$. Other choices
of parameter are possible and $\delta_0$ can take any value,
even much smaller than the resolution of the velocity field
\citep{HernandezCarrasco2011}. This opens many possibilities
that will not be fully explored in this work (see also 
Fig. \ref{fig:lat_mixing} and \ref{ape:smooth}) . Using 
similar parameters for the FSLEs' computation, We also 
investigate the response of the coupled biophysical system to
variable resolution of the velocity field, (see
\cite{HernandezCarrasco2011} for further details about the
sensitivity and robustness of the FSLEs).

The field of FSLEs thus depends on the choice of two length
scales: the initial, $\delta_0$  and the final $\delta_f$
separations. As in previous works \citep{dOvidio2004,
dOvidio2009,HernandezCarrasco2011} we focus on transport
processes at mesoscale, so that $\delta_f$ is taken as about
110 $km$, or 1$\degree$, which is the order of the size of
mesoscale eddies at mid latitudes. To compute $\lambda$ we need
to know the trajectories of the particles, which gives the
Lagrangian character to this quantity. The equations of motion
that describe the horizontal evolution of particle trajectories
in longitudinal and latitudinal spherical coordinates,
$\textbf{x}=(\phi,\theta)$, are:
\begin{eqnarray}
\frac{d\phi}{dt}&=&\frac{u(\phi, \theta, t)}{R \cos {\theta}}, \label{eqsmotiona}\\
\frac{d\theta}{dt}&=&\frac{v(\phi, \theta, t)}{R},
\label{eqsmotionb}
\end{eqnarray}
where $u$ and $v$ represent the eastwards and northwards
components of the surface velocity field, and $R$ is the radius of the Earth
(6371 km).

The ridges of the FSLE field  can be used to define the
Lagrangian Coherent Structures (LCSs)
\citep{Haller2000b,dOvidio2004,dOvidio2009,TewKai2009,HernandezCarrasco2011},
which are useful to characterize the flow from the Lagrangian
point of view \citep{Joseph2002,Koh2002}. Since we are
only interested in the ridges of large FSLE values, the
ones which significantly affect stirring, LCSs can be computed by 
the high values of FSLE which have a line-like
shape. We compute FSLEs by integrating backwards-in-time the
particle trajectories since attracting LCSs (and its associated
unstable manifolds) have a direct physical interpretation
\citep{Joseph2002,dOvidio2004,dOvidio2009}. Tracers, such as
temperature and chlorophyll-a, spread along the attracting
LCSs, thus creating their typical filamental structure
\citep{Lehan2007,Calil2010}.

\subsection{The Biological model}
\label{bio}

The plankton model is similar to the one used in previous
studies by \citet{Oschlies1998,Oschlies1999} and
\citet{Sandulescu2007,Sandulescu2008}. It describes the
interaction of a three-level trophic chain in the mixed layer
of the ocean, including phytoplankton $P$, zoo-plankton $Z$ and
dissolved inorganic nutrient $N$, whose concentrations evolve
in time according to the following equations:

\begin{eqnarray}
\dfrac{dN}{dt}=F_{N}=\Phi _{N}-\beta \dfrac{N}{\kappa_{N}+N}P
+\mu_{N}\left( (1-\gamma) \dfrac{\alpha \eta P^{2}}{\alpha + \eta P^{2}}Z + \mu_{P}P+\mu_{z}Z^{2} \right),
\label{Eq.biolo1}\\
\dfrac{dP}{dt}=F_{P}=\beta \dfrac{N}{\kappa_{N}+N}P-\dfrac{\alpha \eta P^{2}}{\alpha + \eta P^{2}}Z - \mu_{P}P,~~~~~
~~~~~~~~~~~~~~~~~~~~~~~~~~~~~~~~~
\label{Eq.biolo2}\\
\dfrac{dZ}{dt}=F_{Z}=\gamma \dfrac{\alpha \eta P^{2}}{\alpha + \eta P^{2}}Z - \mu_{Z}Z^{2},~~~~~~~~~~~~~~~~~~~~~~~~~~
~~~~~~~~~~~~~~~~~~~~~~~~~~~~
\label{Eq.biolo3}
\end{eqnarray}
where the dynamics of the nutrients, Eq. (\ref{Eq.biolo1}), is
determined by nutrient supply due to the vertical transport $\Phi
_{N}$, its uptake by phytoplankton (2$^{nd}$ term) and its
recycling by bacteria from sinking particles (remineralization) (3$^{rd}$ term).
Vertical mixing which brings subsurface nutrients into
the mixed surface layer of the ocean is parameterized in our coupled
model (see below), since the hydrodynamical part considers only
horizontal 2D transport. The terms in Eq. (\ref{Eq.biolo2}) represent
the phytoplankton growth by consumption of $N$ 
(i.e. primary production $PP =\dfrac{N}{\kappa_{N}+N}P$), the
grazing by zooplankton ($G_{z} =\dfrac{\alpha \eta P^{2}}{\alpha + \eta P^{2}}Z $), 
and natural mortality of phytoplankton. The last
equation, Eq. (\ref{Eq.biolo3}), represents zooplankton growth
by consuming phytoplankton minus zooplankton quadratic
mortality.

\begin{figure}[htb]
\begin{center}
\includegraphics[width=0.70\textwidth, height=0.70\textheight,angle=270]{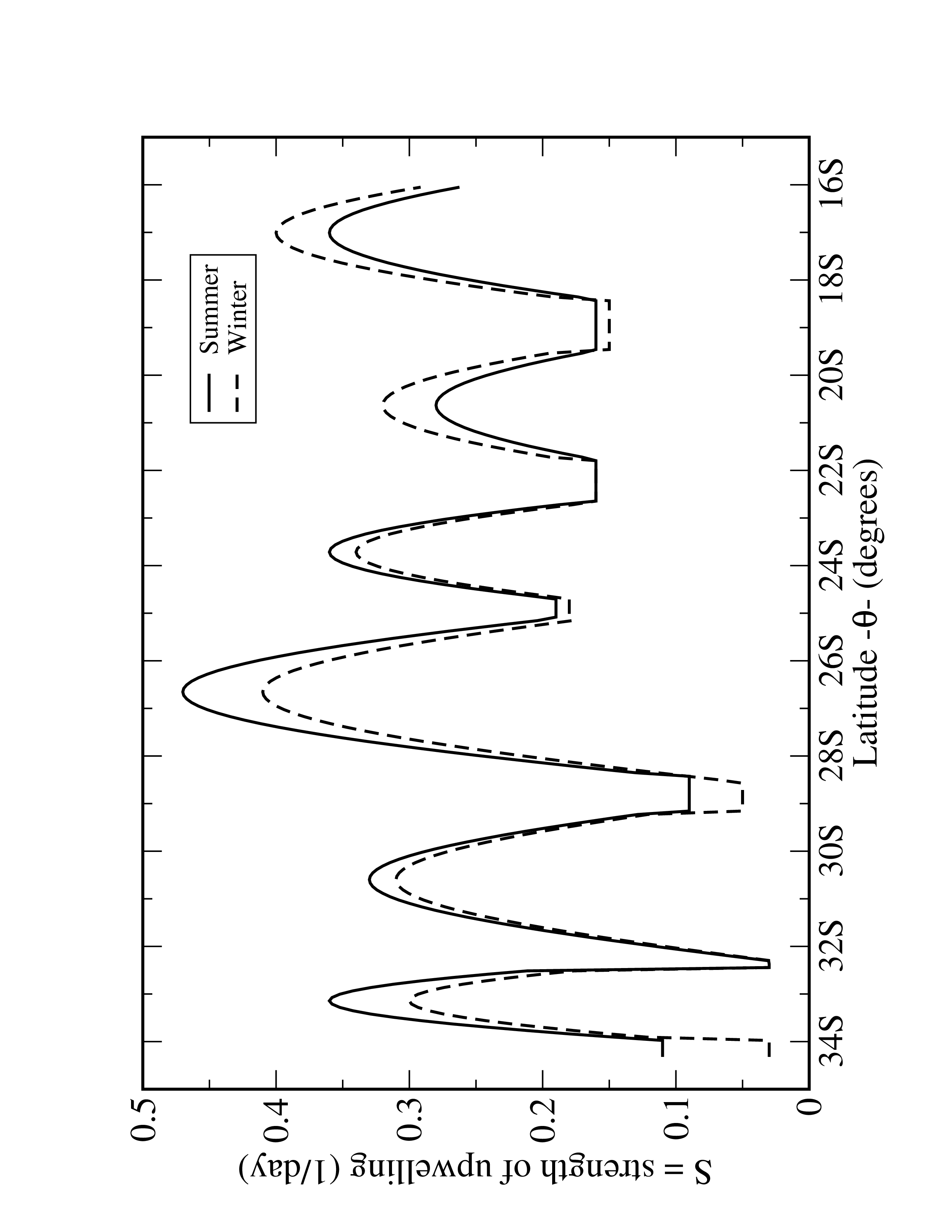}
\end{center}
\caption{Shape and values of the strength ($S$) of the upwelling cells used in the simulations
for winter and summer seasons (following \cite{Veitch2009}).
}
\label{fig:cells}
\end{figure}

An important term of our model is the parameterization 
of the vertical transport of nutrients by coastal upwelling.
Assuming constant nutrient concentration $N_b$ below the mixed
layer, this term can be expressed as:
\begin{equation}
\Phi_{N}(\textbf{x},t)=S(\textbf{x},t)(N_{b}-N(\textbf{x},t)),
\label{eq.vertical_mixing}
\end{equation}
where the function $S$, which depends on time and space (on the two
dimensional location $\textbf{x}$), determines the
amplitude and the spatial distribution of vertical
mixing in the model, thus specifying the strength of the
coastal upwelling. Thus, the function $S$ represents the
vertical transport due to coastal upwelling in our 2D model.
Upwelling intensity along the coast is characterized by a
number of coastal cells of enhanced vertical Ekman driven
transport that are associated with similar fluctuations of the
alongshore wind \citep{Demarcq2003,Veitch2009}. Following these
results, we defined our function $S$ as being null over the
whole domain except in a 0.5$\degree$ wide coastal strip,
varying in intensity depending on the latitude concerned  (see
Fig. \ref{fig:cells}). Six separate upwelling cells, peaking at 
approximately 33$\degree$S, 31$\degree$S, 27.5$\degree$S, 
24.5$\degree$S, 21.5$\degree$S, 17.5$\degree$S,can be
discerned. They are named Cape Peninsula, Columbine/Namaqua, Luderitz, Walvis
Bay, Namibia and Cunene, respectively, Luderitz being the
strongest. For the temporal dependence, $S$ switches between a
summer and a winter parameterization displayed in Fig.
\ref{fig:cells}.

When $\Phi_{N}$ is fixed to either its summer or its winter
shape described in Fig. \ref{fig:cells}, the dynamical system
given by Eqs. (\ref{Eq.biolo1},\ref{Eq.biolo2},\ref{Eq.biolo3})
evolves towards an equilibrium distribution for $N$, $P$ and
$Z$. The transient time to reach equilibrium is typically $60$
days with the initial concentrations used (see Sec.
\ref{coupling}). The parameters are set following a study by
\cite{Pasquero2004} and are listed in Table \ref{tab.bio}.

\begin{table}[ht]
\begin{center}
\begin{tabular}{l l l}
\hline
Parameter & Symbol~~ & ~~~~~Value\\
\hline
Phytoplankton growth rate&  ~~~$\beta$ & ~~0.66 day$^{-1}$\\
Prey capture rate&  ~~~$\eta$ & ~~1.0 (mmol N m$^{-3}$)$^{-2}$ day$^{-1}$\\
Assimilation efficiency of Zooplankton&  ~~~$\gamma$ & ~~~0.75\\
Maximum grazing rate &  ~~~$a$  &  ~~2.0 day$^{-1}$\\
Half-saturation constant for N uptake&  ~~~$k_{N}$ & ~~0.5 mmol N m$^{-3}$\\
Inefficiency of remineralization&  ~~~$\mu_{N}$ & ~~0.2\\
Specific mortality rate&  ~~~$\mu_{P}$ &  ~~0.03 day$^{-1}$\\
(Quadratic) mortality&  ~~~$\mu_{Z}$ &  ~~0.2 (mmol N m$^{-3}$)$^{-1}$ day$^{-1}$\\
Nutrient concentration bellow mixed layer&  ~~~$N_{b}$ &  ~~8.0 mmol N m$^{-3}$\\
\hline
\end{tabular}
\end{center}
\caption{List of parameters used in the biological model.
 }
\label{tab.bio}
\end{table}

\subsection{Coupling hydrodynamical and biological models in Benguela.}
\label{coupling}

We used the velocity fields provided by \citep{Sudre2008} 
and \citep{Gutknecht2013} to do offline coupling with the NPZ model. 
The evolution of simulated concentrations advected within a flow is 
determined by the coupling between the hydrodynamical and biological
models, as described by an advection-reaction-diffusion system.
The complete model is given by the following system of partial
differential equations:

\begin{eqnarray}
\dfrac{\partial N}{\partial t}+\textbf{v}\cdot \nabla N=F_{N}+D\nabla^{2}N,\label{coupledsystem1}\\
\dfrac{\partial P}{\partial t}+\textbf{v}\cdot \nabla P=F_{P}+D\nabla^{2}P,\label{coupledsystem2}\\
\dfrac{\partial Z}{\partial t}+\textbf{v}\cdot \nabla Z=F_{Z}+D\nabla^{2}Z.\label{coupledsystem3}
\end{eqnarray}

The biological model is the one described previously by the
functions $F_N$, $F_P$ and $F_Z$. Horizontal advection
is the 2D velocity field $\textbf{v}$, which is obtained from satellite
data or from the ROMS model.
We add also an eddy diffusion term, via the $\nabla^{2}$ operator,
acting on $N$, $P$, and $Z$
to incorporate the unresolved small-scales
which are not explicitly taken into account by the velocity
fields used.

The eddy diffusion coefficient, $D$, is given by Okubo's formula \citep{Okubo1971},
$D(l)=2.055 * 10^{-4}~l^{1.15}$, where $l$ is the value of the resolution, in meters,
corresponding to the angular resolution $l=\Delta_0$.
The formula gives the values
$D$=26.73~ $m^{2}/s$ for \textit{Satellite1/4} and
\textit{ROMS1/4}, and $D$=7.4$~ m^{2}/s$ for \textit{ROMS1/12}.

The coupled system Eqs.
(\ref{coupledsystem1}),(\ref{coupledsystem2}) and
(\ref{coupledsystem3}) is solved numerically by the
semi-Lagrangian algorithm described in \cite{Sandulescu2007},
combining Eulerian and Lagrangian schemes. The initial
concentrations of the tracers were taken from \cite{Kone2005}
and they are $N_{0}=1\ mmol N m^{-3}$ , $P_{0}=0.1\ mmol N
m^{-3}$, and $Z_{0}=0.06 \ mmol N m^{-3}$. The inflow
conditions at the boundaries are specified in the following
way: at the eastern corner, and at the western and southern edges of the
computational domain fluid parcels enter with very low
concentrations  ($N_{L}=0.01N_{0}\ mmol N m^{-3}$,
$P_{L}=0.01P_{0}\ mmol N m^{-3}$, and $Z_{L}=0.01Z_{0 }\ mmol N
m^{-3}$). Across the northern boundary, fluid parcels enter with
higher concentrations ($N_{H}=5 \ mmol N m^{-3}$, $P_{H}=0.1 \
mmol N m^{-3}$, and $Z_{H}=0.06 \ mmol N m^{-3}$). Nitrate 
concentrations are derived from CARS climatology \citep{Condie2006}, 
while P and Z concentrations are taken from \cite{Kone2005}. 
The integration time step is $dt=6$ hours.

To convert the modeled $P$ values, originally in $mmol N.m^{-}3$, 
into $mg~m^{-3}$ of chlorophyll, we used a standard ratio of 
$Chloro/Nitrogen = 1.59$ as prescribed by \cite{Hurtt1996} and \cite{Doney1996}.
In the following we refer to as ``simulated chlorophyll'' for the concentrations 
derived from the simulated phytoplankton P, after applying the conversion 
ratio (see above); and we use ``observed chlorophyll'' for the 
chlorophyll-a measured by SeaWIFS. 

\section{Results and discussion}
\label{sec:results}

\subsection{Validation of our simple 2D idealized setting using satellite data}
\subsubsection{Horizontal stirring}
\label{FSLE}

\begin{figure}[htb]
\begin{center}
\includegraphics[width=0.99\textwidth, height=0.65\textheight]{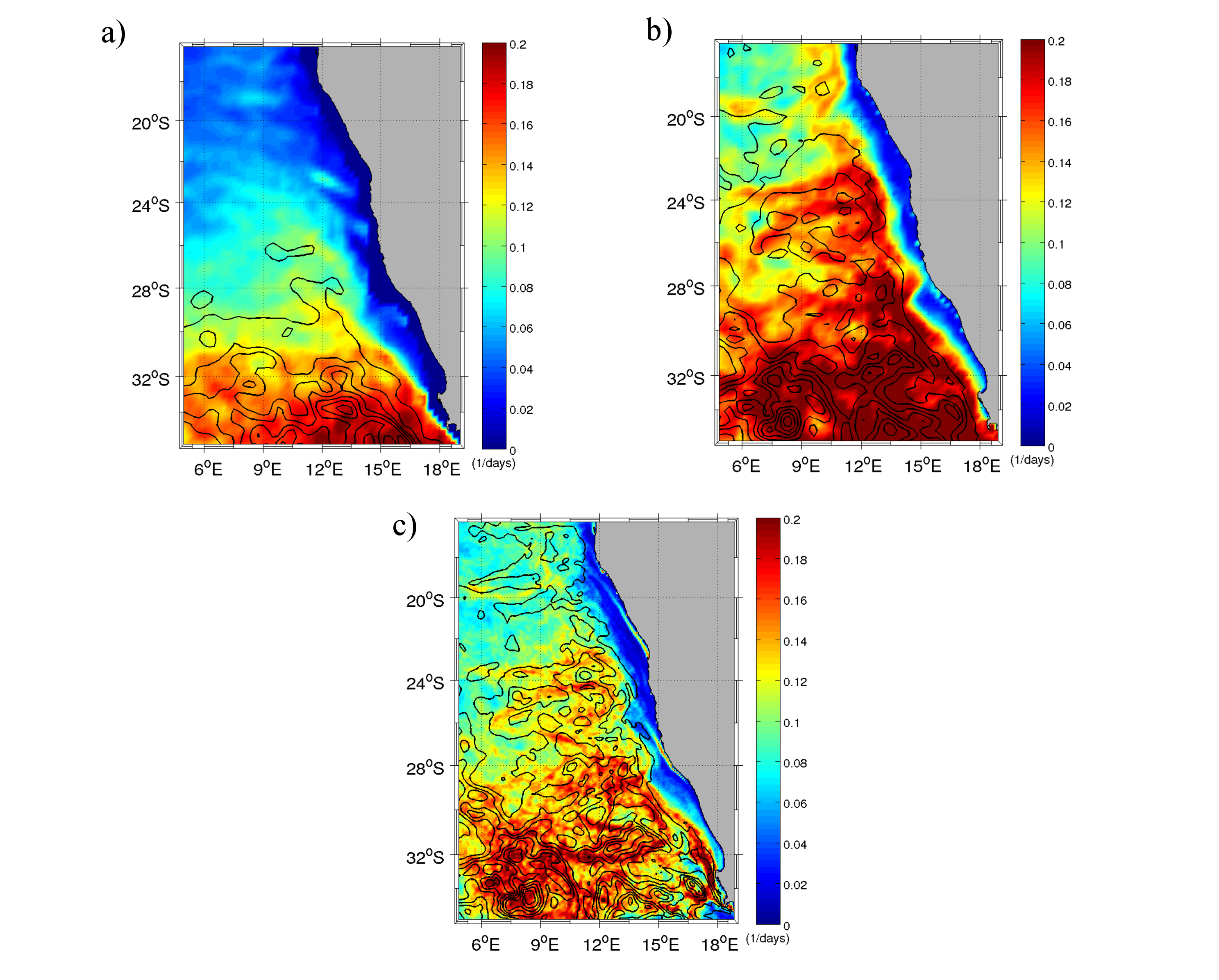}
\end{center}
\caption{Spatial distribution of time average
of weekly FSLE maps in the Benguela region. a) Three years average
using data set \textit{Satellite1/4};  b) one year average using \textit{ROMS1/4};
c) one year average using
\textit{ROMS1/12}. The units of the colorbar are $1/days$. The black lines are contours of
annual EKE. The separation between contour levels is 100$cm^{2}/s^{2}$.
 }
\label{fig:mixing}
\end{figure}

We compute the FSLE with an initial separation of particles
equal to the spatial resolution of each velocity field
($\delta_0$= 1/4$\degree$ for \textit{Satellite1/4} and
\textit{ROMS1/4}, and $\delta_0$= 1/12$\degree$ for
\textit{ROMS1/12}), an a final distance of $\delta_f$=
1$\degree$ to focus on transport processes by mesoscale
structures at mid latitudes. The areas of more intense
horizontal stirring due to mesoscale activity can be identified
by large values of temporal averages of backward FSLEs (see
Figure \ref{fig:mixing}). 
While there are visible differences between
the results from the different velocity fields, especially in
the small-scale patterns, the spatial pattern are quantitatively well
reproduced. For instance, spatial correlation coefficient $R^{2}$
between FSLEs map from \textit{Satellite1/4} and from
\textit{ROMS1/4} is $0.81$. Correlation coefficients between
\textit{Satellite1/4} and \textit{ROMS1/12} on one hand, and
between \textit{ROMS1/4} and \textit{ROMS1/12} on the other
hand, are lower ($0.61$ and $0.77$ respectively) since the FSLE
were computed on a different resolution. More details on the
effect on the grid resolution when computing FSLEs can be found
in \citet{HernandezCarrasco2011}. For all datasets, high
stirring values are observed in the southern region, while the
northern area displays significantly lower values, in line with
\cite{Rossi2009}.  Note that the separation is well marked for
\textit{Satellite1/4} where high and low values of FSLE occur
below and above a line at $27\degree$ approximately. In the
case of ROMS flow fields, the stirring activity is more
homogeneously distributed, although the north-south gradient is
still present. We associate this latitudinal gradient with the
injection of energetic Agulhas rings, the intense jet/bathymetry interactions 
and with other source of flow instabilities in the southern Benguela. 
Following \cite{Gruber2011} we compute the EKE, another proxy of 
the intensity of mesoscale activity. There are regions with distinct
dynamical characteristics as the southern subsystem is
characterized by larger EKE values than the northern area,
in good agreement with the analysis arising from FSLEs
(Fig. \ref{fig:mixing}). Spatial correlations (not
shown) indicate that EKE and FSLE patterns are well correlated
using a non-linear fitting (power law). For instance, EKE
and FSLE computed on the velocity field from
\textit{Satellite1/4} exhibit a $R^{2}$ of $0.86$ for the
non-linear fitting: $FSLE=0.009\cdot EKE^{0.49}$. This is in
agreement with the initial results from
\citet{Waugh2006,Waugh2008}, for a related dispersion
measurement, and confirmed for the Benguela region by the 
thorough investigations of EKE/FSLE relationship by
\citet{HernandezCarrasco2011b}.

\begin{figure}[htb]
\begin{center}
\includegraphics[width=0.50\textwidth, height=0.31\textheight]{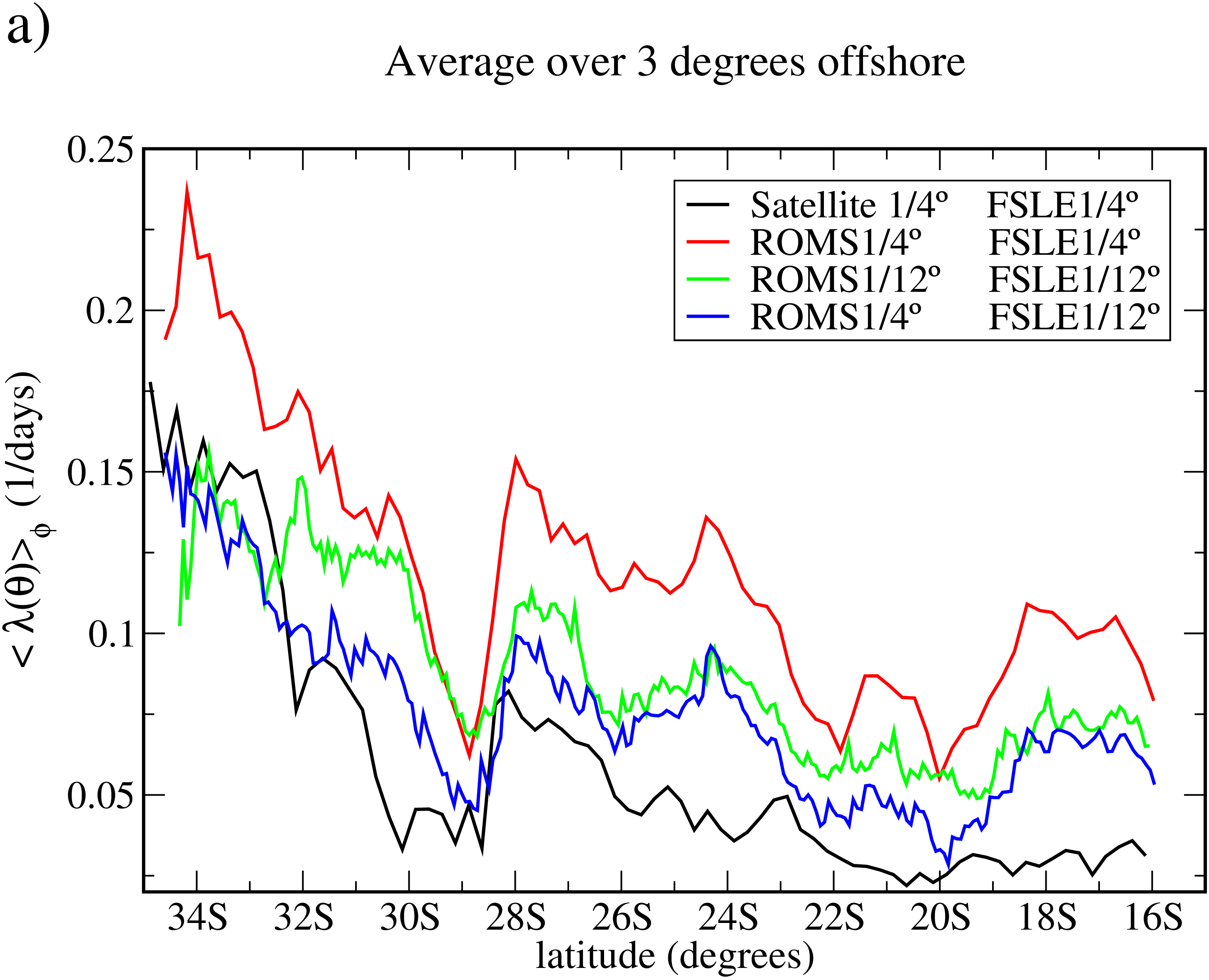}\\
\hspace{0.78cm}
\includegraphics[width=0.43\textwidth, height=0.43\textheight, angle=270]{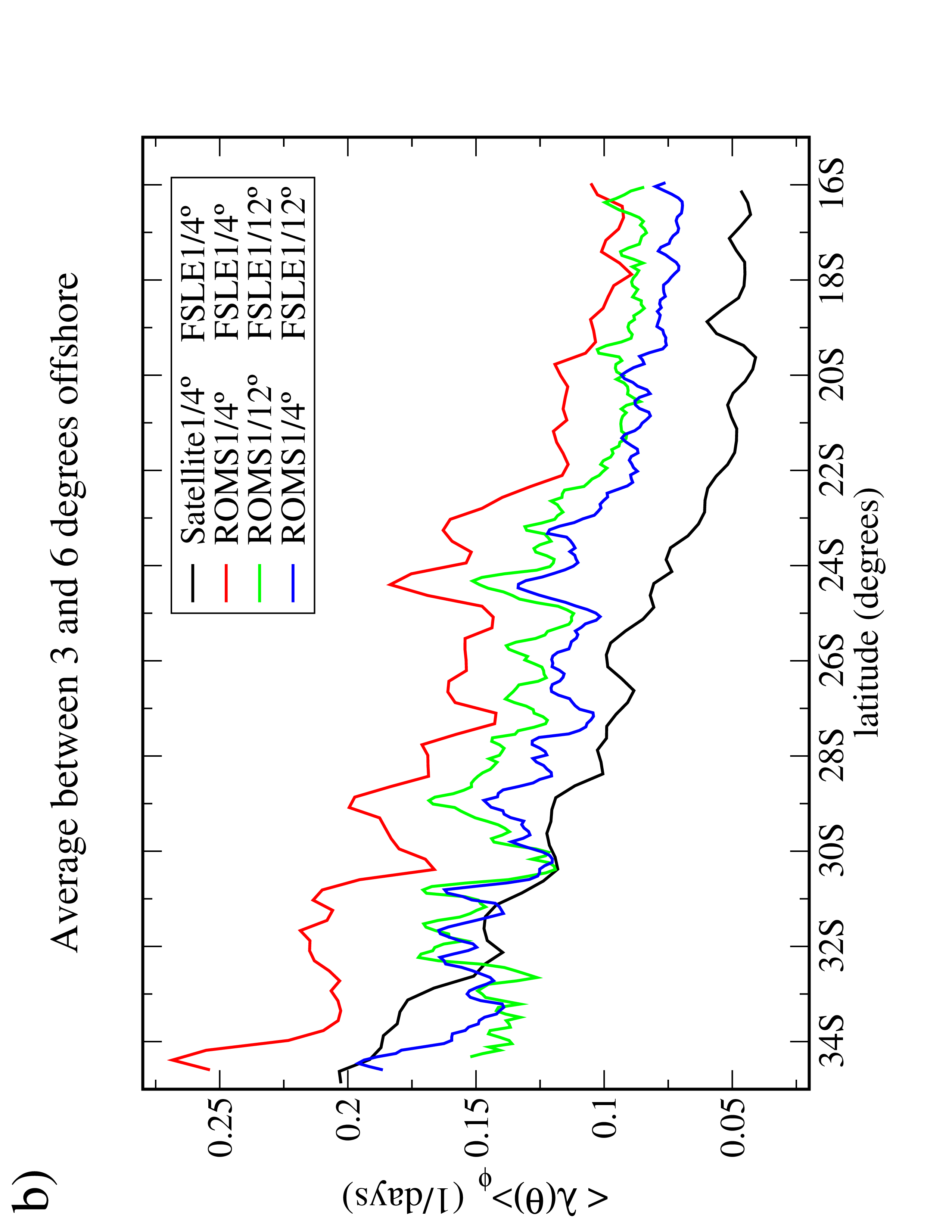}
\end{center}
\caption{Zonal average on coastal bands of the FSLE time averages from
Fig. \ref{fig:mixing} as a function of latitude. a) from the coast to 3 degrees offshore;
 b) between 3 and 6 degrees offshore.
}
\label{fig:lat_mixing}
\end{figure}

To analyze the variability of horizontal mixing with latitude,
we compute longitudinal averages of the plots in Fig.
\ref{fig:mixing} for two different coastally-oriented strips
extended: a) from the coast to $3\degree$ offshore, and b) from
$3\degree$ to $6\degree$ offshore (see Fig.
\ref{fig:lat_mixing}). It allows analyzing separately subareas
characterized by distinct bio-physical characteristics 
(see also \cite{Rossi2009}), the
coastal upwelling ($3 \degree$ strip) with high plankton
biomasses and moderated mesoscale activity, and the open ocean
(from $3\degree$ to $6\degree$ offshore) with moderated
plankton biomasses and high mesocale activity. It is clear that
horizontal stirring decreases with decreasing latitude. 
In Fig. \ref{fig:lat_mixing} (a) we see that, for
\textit{Satellite1/4}, the values of FSLEs decay from $0.18 \
days^{-1}$ in the southern to $0.03\ days^{-1}$ in the northern
area, with similar significant decays for \textit{ROMS1/4} and
\textit{ROMS1/12}. Specifically the North-South difference for
\textit{Satellite1/4}, \textit{ROMS1/4} and \textit{ROMS1/12}
are of the order of $0.15 \ days^{-1}$ , $0.15\ days^{-1}$ and
$0.08\ days^{-1}$, respectively, confirming a lower latitudinal
gradient for the case of \textit{ROMS1/12}. 

Note that there
are differences in the stirring values (FSLEs) depending on the
type of data, their resolution, the averaging strip, and the
grid size of FSLE computation. In general, considering
velocities with the same resolution, the lower values
correspond to \textit{Satellite1/4} as compared to
\textit{ROMS1/4}. On average, values of stirring from
\textit{ROMS1/4} are larger than those from \textit{ROMS1/12},
whereas we would expect the opposite considering the higher
resolution of the latter simulation favoring small scales
processes. However, this comparison is hampered by the fact 
that spatial means of FSLE values are reduced when
computing them on grids of higher resolution, because the
largest values become increasingly concentrated in thinner
lines, a consequence of their multifractal character
\citep{HernandezCarrasco2011}. Indeed, one 
can not compare consistently two FSLEs field computed 
on a different resolution, whatever the intrinsic resolution 
of the velocity field is. The FSLEs computed on a 1/4$\degree$ grid 
(black and red lines on Fig. \ref{fig:lat_mixing}) cannot be directly compared to 
FSLE fields computed on a 1/12$\degree$ grid (green line Fig. \ref{fig:lat_mixing}) 
(see \cite{HernandezCarrasco2011}).
Note however that when FSLEs are computed using the \textit{ROMS1/12} 
and \textit{ROMS1/4} flows but on the same FSLE grid with a fixed 
resolution of 1/12$\degree$, one finds smaller values of FSLEs 
for the coarser velocity field (\textit{ROMS1/4}) (see green and blue lines in Fig.
\ref{fig:lat_mixing}). The effect of reducing the velocity
spatial resolution on the FSLE calculations is considered more
systematically in \ref{ape:smooth}. FSLE values 
obtained from the same FSLE-grid increase as the resolution of the 
velocity-grid becomes finer (Fig. \ref{fig.comparison_FSLEsmooth})
A general observation consistent between all datasets is that 
horizontal mixing is slightly less intense and more variable
in the region of coastal upwelling (from
the coast to 3$\degree$ offshore) than within the transitional
area with the open ocean (3-6$\degree$ offshore). Note also
that a low-stirring region is observed within the 3$\degree$
width coastal strip from $28\degree$ to $30\degree$S on all
calculations. These observations confirm that the ROMS model is
representing well the latitudinal variability of the stirring
as measured from FSLE based on satellite data. These
preliminary results indicate that Lyapunov exponents and
methods could be used as a diagnostic to validate the
representation of mesoscale activity in eddy-resolving oceanic
models, as suggested recently by \citet{Titaud2011}.
Overall, the variability of stirring activity in the Benguela 
derived from the simulated flow fields is in good agreement 
with the satellite observations.

\subsubsection{Simulated phytoplankton concentrations}
\label{evo_P}

\begin{figure}[htb]
\begin{center}
\includegraphics[width=0.99\textwidth, height=0.65\textheight]{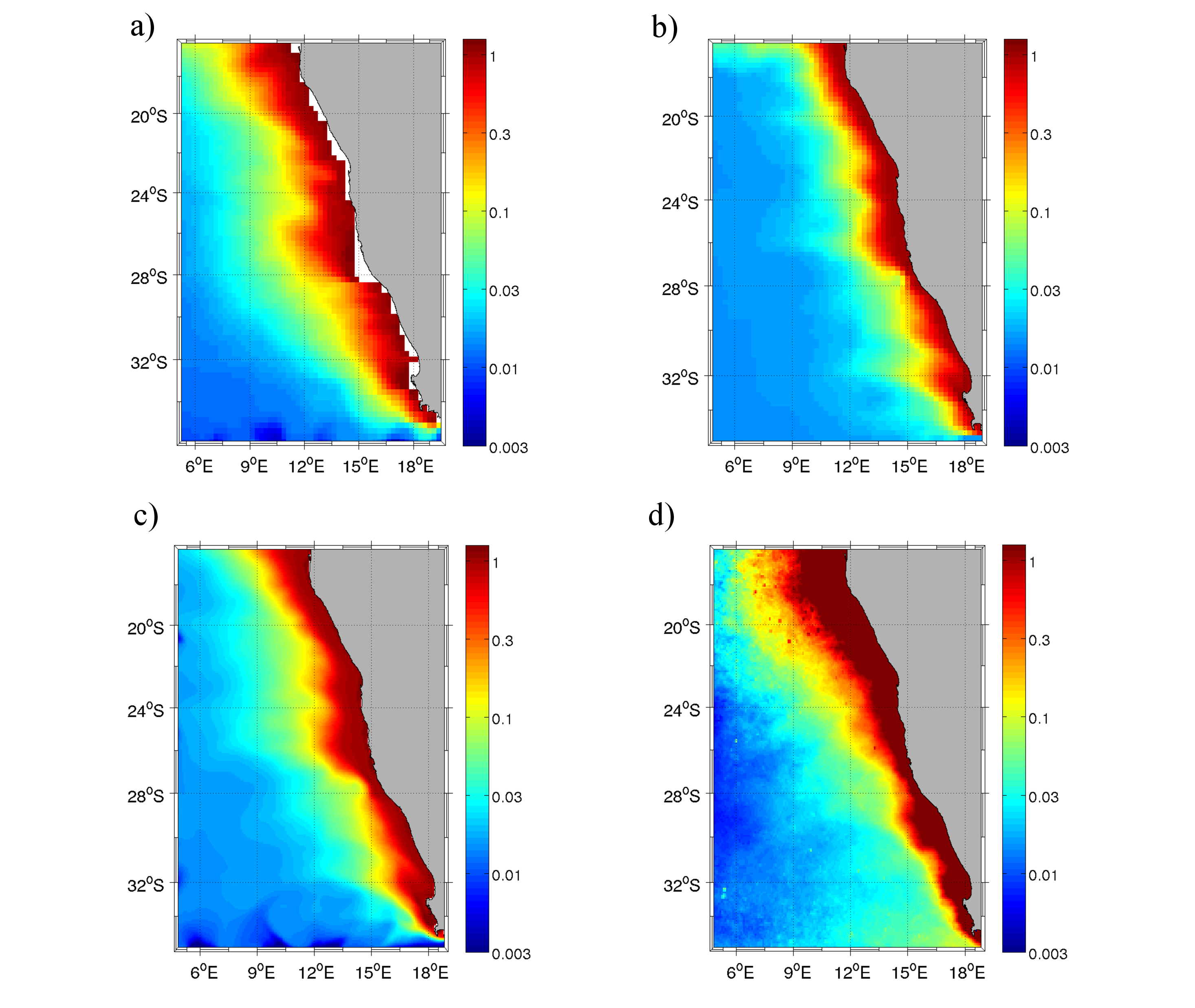}
\end{center}
\caption{
Spatial distribution of:
a) Three years average of simulated chlorophyll using \textit{Satellite1/4},
b) One year average of simulated chlorophyll using \textit{ROMS1/4}, 
c) Same than b) but using \textit{ROMS1/12}, d) Three years average of observed 
chlorophyll derived from monthly SeaWIFS data.
The units of the colorbar are $mg/m^{3}$. Logarithmic scale 
is used to improve the visualization of 
gradients in nearshore area.
}
\label{fig:phyto_average}
\end{figure}

Evolution of $N$, $P$ and $Z$ over space and time is obtained
by integrating the systems described by Eqs.
(\ref{coupledsystem1}), (\ref{coupledsystem2}) and
(\ref{coupledsystem3}). The biological model is coupled to the
velocity field after the spin-up time needed to reach stability
($60$ days). Analysing the temporal average of simulated chlorophyll
(Fig. \ref{fig:phyto_average}), we found that coastal regions
with high $P$ extend approximately, depending on latitude,
between half a degree and two degrees offshore. It is
comparable with the pattern obtained from the satellite-derived
chlorophyll data (Fig.\ref{fig:phyto_average} d)).  
The spatial
correlation between averaged simulated and satellite
chlorophyll is as follows: $R^{2} = 0.85$ for
\textit{Satellite1/4} versus \textit{SeaWIFS}; $R^{2} = 0.89$
for \textit{ROMS1/4} versus \textit{SeaWIFS} and $R^{2} = 0.85$
for \textit{ROMS1/12} versus \textit{SeaWIFS}. Despite the very
simple setting of our model, including the parameterization of
the coastal upwelling, the distribution of phytoplankton
biomass is relatively well simulated in the Benguela area. Note
however that our simulated chlorophyll values are about
$\simeq$ 3-4 times lower than satellite data. Many biological and 
physical factors not taken into account in this simple setting could 
be invoked to explain this offset. Another possible explanation is 
the low reliability of ocean color data in the optically complex
coastal waters \citep{Melin2007}.

\begin{figure}[htb]
\begin{center}
\includegraphics[width=0.90\textwidth, height=0.55\textheight]{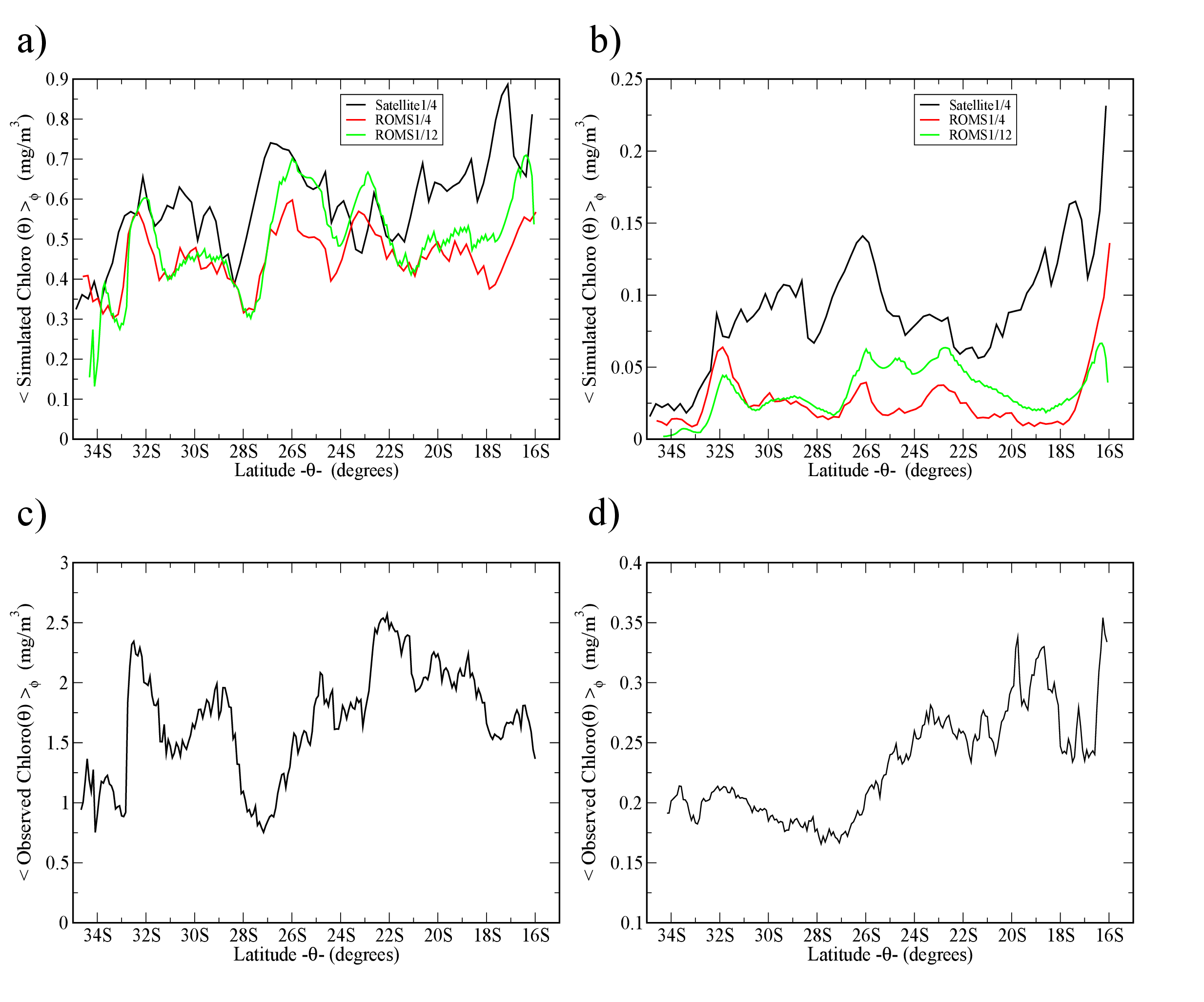}
\end{center}
\caption{
Zonal mean of simulated chlorophyll on a coastally oriented strip from the
coast to 3 degrees (a) and from 3 degrees to 6 degrees offshore (b), plotted as a
function of latitude. Zonal average of observed chlorophyll (SeaWIFS) over 
a coastal band from the coast to 3 degrees (c) and from 3 degrees
to 6 degrees offshore (d).
}
\label{fig:phyto_lat}
\end{figure}

We now examine the latitudinal distribution of $P$ comparing
the outputs of the numerical simulations versus the satellite
chlorophyll-a over different coastally oriented strips
(Fig.\ref{fig:phyto_lat}). Simulated $P$ concentrations are
higher in the northern than in the southern area of Benguela,
in good agreement with the chlorophyll-a data derived from
satellite. A common feature is the minimum located just below
the Luderitz upwelling cell (28$\degree$S), which may be
related to the presence of a physical boundary, already studied
and named the LUCORC barrier by \cite{Shannon2006} and
\cite{Lett2007}. The decrease of $P$ concentration is clearly
visible in the open ocean region of the \textit{Satellite1/4}
case (Fig. \ref{fig:phyto_lat} b)). Correlations of zonal
averages between simulated and satellite chlorophyll-a are poor
when considering the whole area ($R^{2}$ ranging from 0.1 to
0.5). However, when considering each subsystem (northern and
southern) independently, high correlation coefficients are
found for the south Benguela ($R^{2}$ around 0.75), but not for
the north. This indicates that our simple modelling approach is
able to simulate the spatial patterns of chlorophyll in the
south Benguela, but not properly in the northern part. In the
north, other factors not considered here (such as the 3D flow,
the varying shelf width, the external inputs of nutrients,
realistic non-climatologic forcings, complex biogeochemical
processes, etc...) seem to play an important role in
determining the surface chlorophyll-a observed from space.

\subsection{Relationship between phytoplankton and horizontal stirring.}

\begin{figure}[htb]
\begin{center}
\includegraphics[width=0.99\textwidth, height=0.99\textheight]{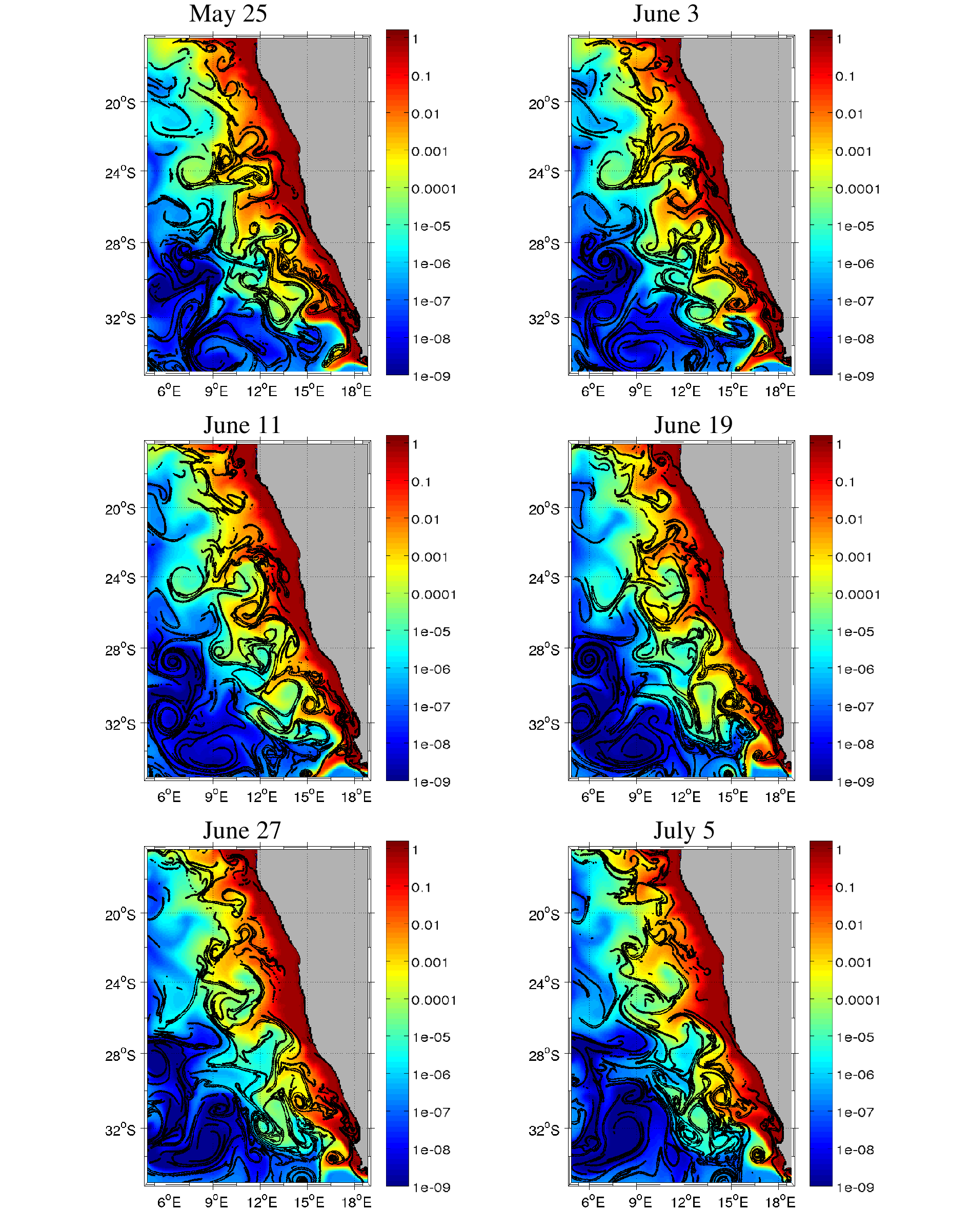}
\end{center}
\caption{Snapshots every $8$ days of large (top $30 \%$) values of FSLE superimposed on
simulated chlorophyll concentrations calculated from \textit{ROMS1/12} in $mg/m^{3}$. Logarithmic scale for
chlorophyll concentrations is used to improve the visualization of the structures
}
\label{fig:evo_eddie}
\end{figure}

In Fig. \ref{fig:evo_eddie} we show six selected snapshots of 
chlorophyll concentrations every $8$ days
during a $32$ days period for  \textit{ROMS1/12}.
Since both ROMS simulations were climatologically forced runs, the dates do
not correspond to a specific year. The most relevant feature is the larger value of
concentrations near the coast due to the injection of nutrients.
Obviously the spatial distribution of $P$ is strongly influenced by
 the submeso- and meso-scale structures such as
filaments and eddies, especially in the southern subsystem.
Differences are however observed between the three data sets.
In particular, it seems that for \textit{Satellite1/4} and
\textit{ROMS1/12} the concentrations extend further offshore
than for \textit{ROMS1/4} (not shown). In \ref{ape:smooth} we
provide additional analysis of the effect of the velocity
spatial resolution on phytoplankton evolution. We found
that velocity data with different resolution produces similar
phytoplankton patterns but larger absolute values of
concentrations as the spatial resolution of the velocity field
is refined (see \cite{Mahadevan2000,Levy2001}), supporting the
need to compare different spatial resolutions.


Several studies \citep{Lehan2007,dOvidio2009,Calil2010} have
shown that transport of chlorophyll distributions in the marine
surface is linked to the motion of local maxima or ridges of
the FSLEs. This is also observed in our numerical setting when
superimposing contours of high values of FSLE (locating the
LCSs) on top of phytoplankton concentrations for
\textit{ROMS1/12} (see Fig. \ref{fig:evo_eddie}). In some
regions $P$ concentrations are constrained and stirred by lines
of FSLE. For instance, the elliptic eddy-like structure at $13\
\degree$E, $32\ \degree$S is characterized by high phytoplankton 
concentrations at its edge, but relatively low in its core. 
This reflects the fact
that tracers, even active such as chlorophyll, still disperse
along the LCSs.

From Fig. \ref{fig:phyto_lat} it is clear that phytoplankton
biomass has a general tendency to decrease with latitude, an
opposite tendency to the one exhibited by stirring (as inferred
from the FSLEs and EKE distributions in Figs. \ref{fig:mixing}
and \ref{fig:lat_mixing}) for the three data sets. Moreover,
note that the minimum of phytoplankton located just below the
LUCORC barrier at $28 \degree$S (Fig. \ref{fig:phyto_lat})
coincides with a local maximum of
stirring that might be responsible for this barrier (Fig.
\ref{fig:lat_mixing} a). Spatial mean and latitudinal
variations of FSLE and chlorophyll-a analyzed together suggest
an inverse relationship between those two variables. 
The 2D vigorous stirring in the south
and its associated offshore export seem sufficient to simulate
reasonably well the latitudinal patterns of $P$. The numerous
eddies released from the Agulhas system and generally travelling north-westward, 
associated with the elevated mesoscale activity in the south Benguela, might 
inhibit the development of $P$ and export unused nutrients 
toward the open ocean. Although \citet{Gruber2011} invoked the 
offshore subduction of unused nutrients (3D effect), our results suggest that 2D 
offshore advection and intense horizontal mixing could by themselves
affect negatively the phytoplankton growth in the southern 
Benguela.

\begin{figure}[htb]
\begin{center}
\includegraphics[width=0.40\textwidth, height=0.35\textheight, angle=270]{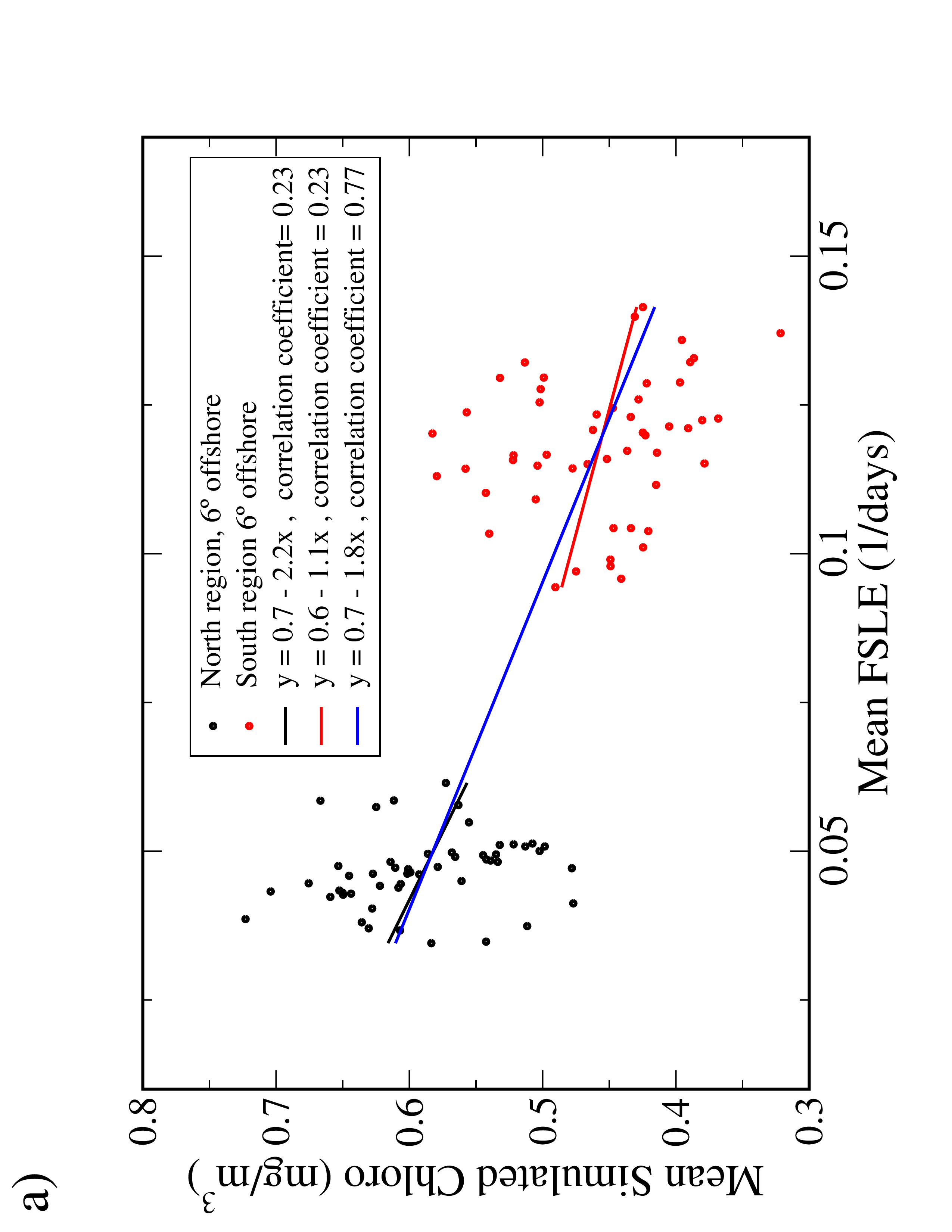}
\includegraphics[width=0.40\textwidth, height=0.35\textheight, angle=270]{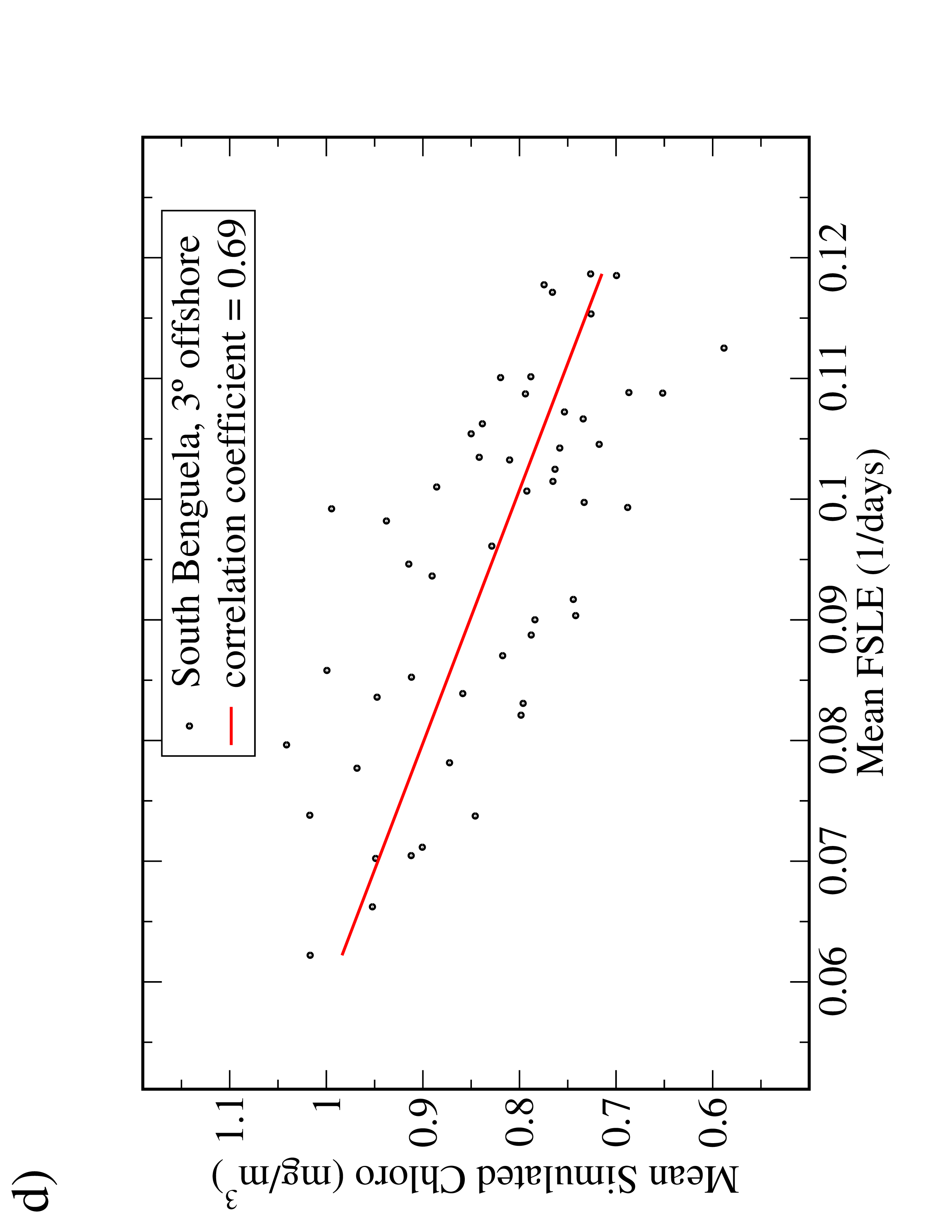}
\includegraphics[width=0.40\textwidth, height=0.35\textheight, angle=270]{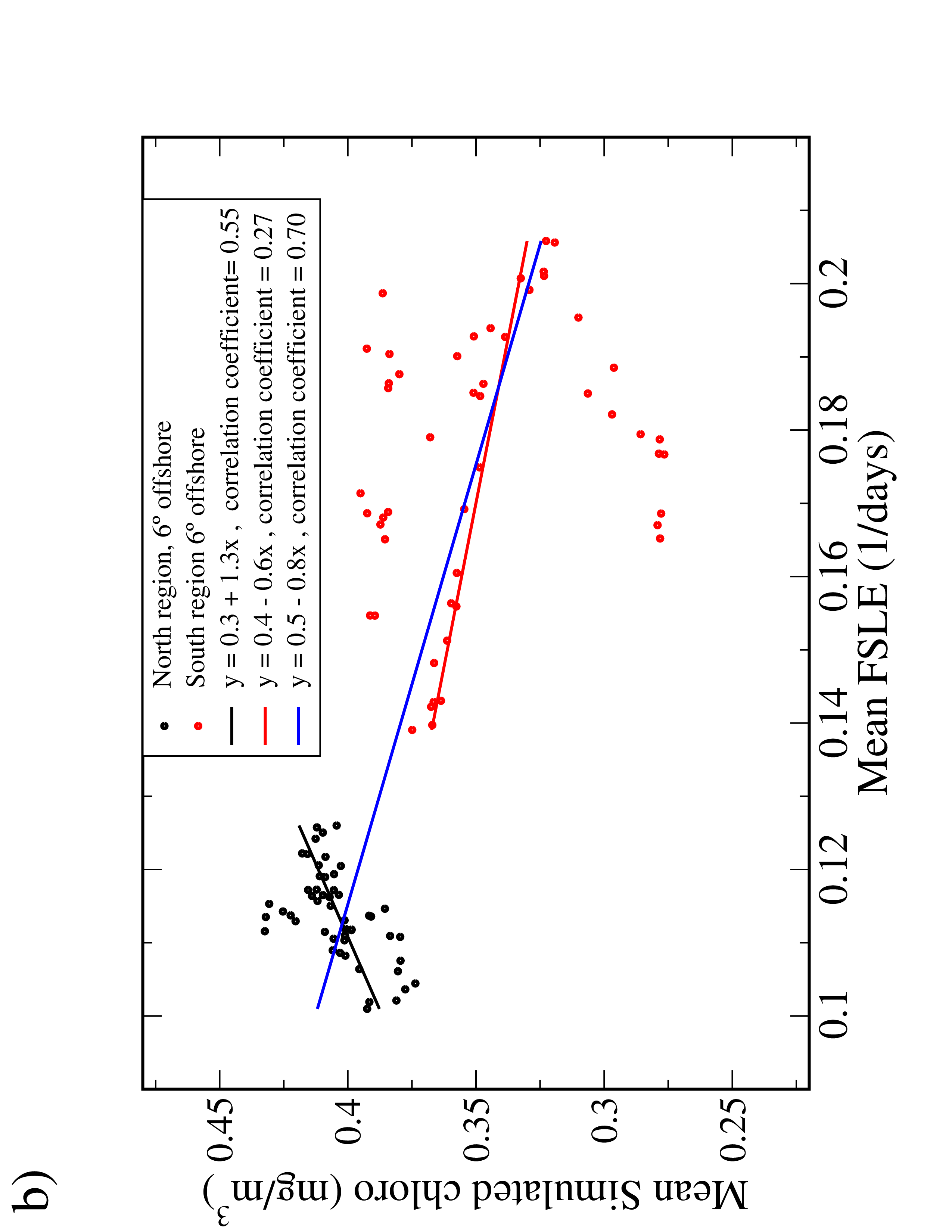}
\includegraphics[width=0.40\textwidth, height=0.35\textheight, angle=270]{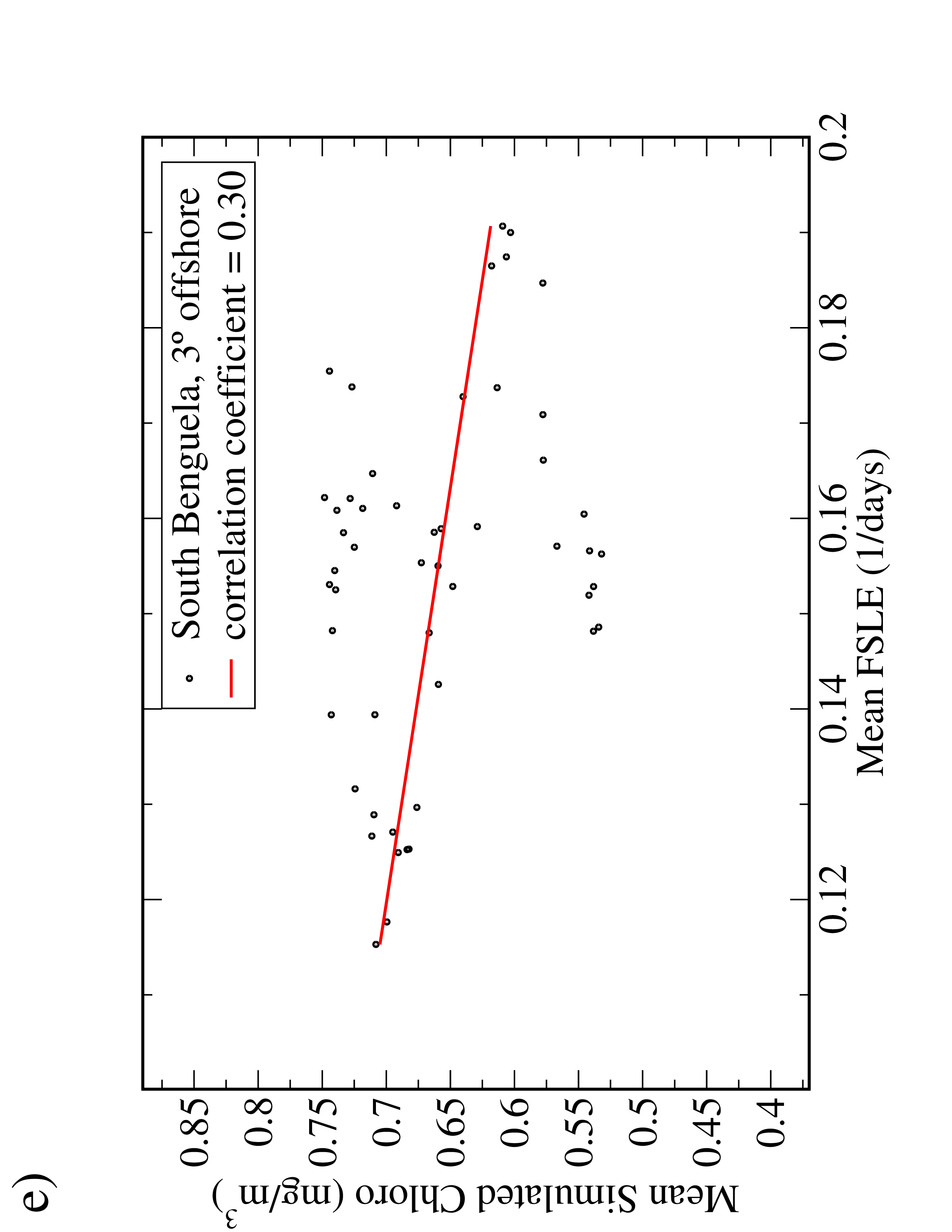}
\includegraphics[width=0.40\textwidth, height=0.35\textheight, angle=270]{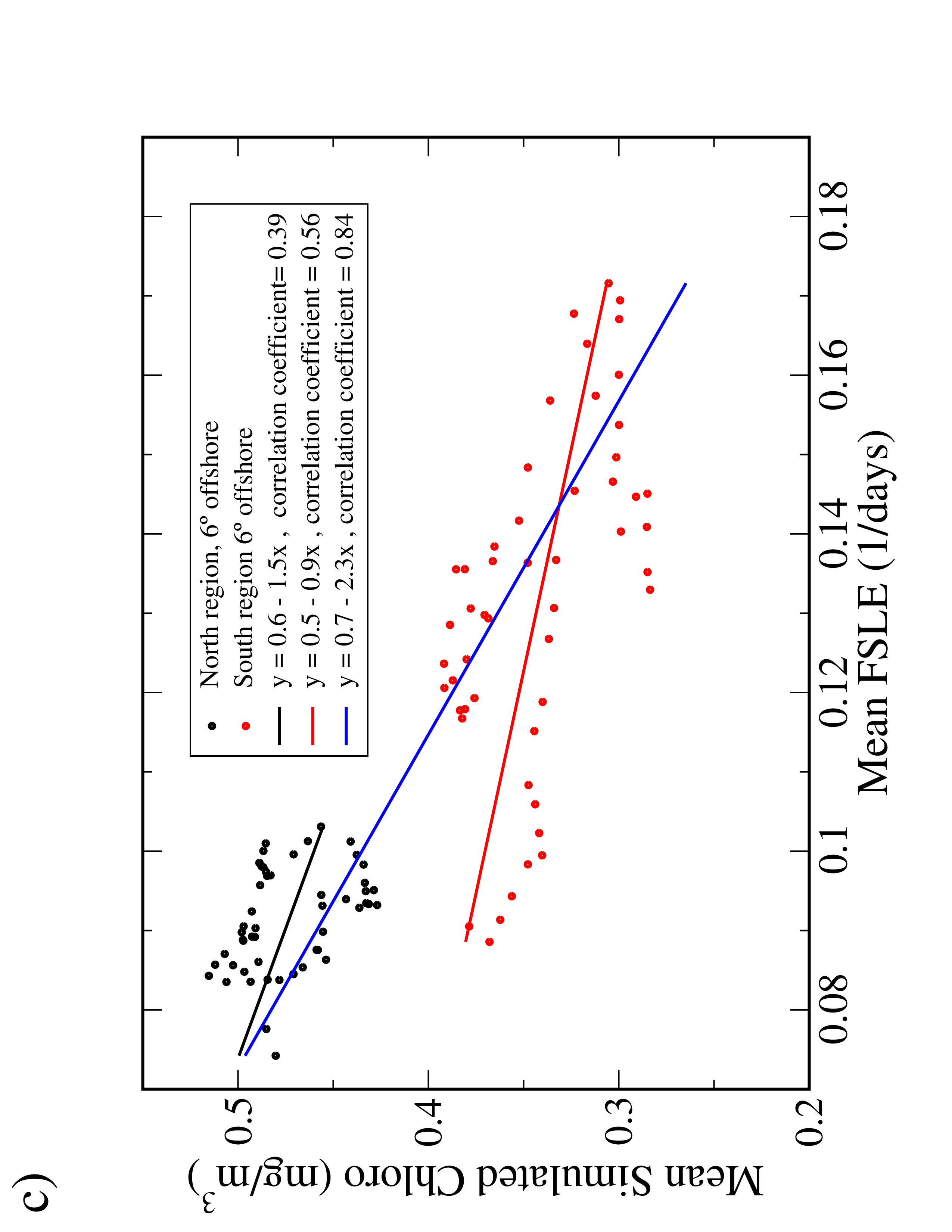}
\includegraphics[width=0.40\textwidth, height=0.35\textheight, angle=270]{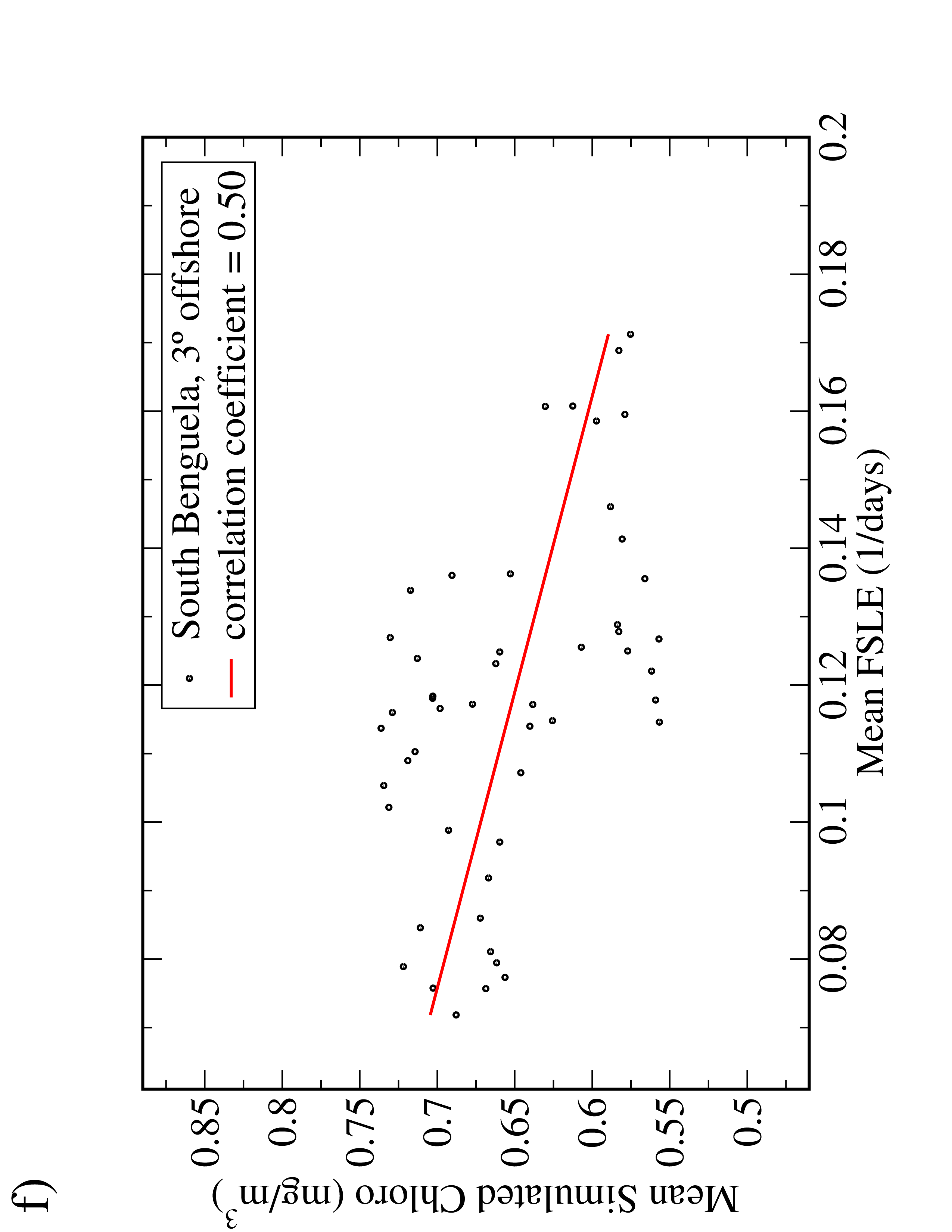}
\end{center}
\caption{ Weekly values of spatial averages of simulated chlorophyll versus
weekly values of spatial averages of FSLE,
where the average are over the whole area (6 $\degree$  from the coast)
and in North and South subareas of Benguela. a) \textit{Satellite1/4},
b) \textit{ROMS1/4} and c) \textit{ROMS1/12}. Right column plots
the average over 3$\degree$ offshore in the south region: d) \textit{Satellite1/4},
e) \textit{ROMS1/4} and f) \textit{ROMS1/12}
 }
\label{fig:Phyto_FSLE}
\end{figure}

To study quantatively the negative effect of horizontal
stirring on phytoplankton concentration, we examine the correlation between the spatial
averages -- over each subregion (North and South) and the whole
area of study -- of every weekly map of FSLE and the spatial
average of the corresponding weekly map of simulated $P$,
considering each of the three velocity fields
(Fig.\ref{fig:Phyto_FSLE}). For all cases, a negative
correlation between FSLEs and chlorophyll emerges: the
higher the surface stirring/mixing, the lower the biomass
concentration. The correlation coefficient taking into account
the whole area is quite high for all the plots, $R^{2}$=0.77
for \textit{Satellite1/4}, 0.70 for \textit{ROMS1/4} and 0.84
for \textit{ROMS1/12} , and the slopes (blue lines in
Fig.\ref{fig:Phyto_FSLE} have the following values: -1.8 for
\textit{Satellite1/4}, -0.8 for \textit{ROMS1/4} and -2.3 for
\textit{ROMS1/12}. The strongest negative correlation is found 
for the setting with \textit{ROMS1/12}. Note that,
similarly to the results of \cite{Rossi2008,Rossi2009} and
\cite{Gruber2011}, the negative slope is larger but less robust
when considering the whole area rather than within every
subregion. Moreover, if we average over the coastal strip (from
coast to 3$\degree$ offshore) and only in the south region
(Fig.\ref{fig:Phyto_FSLE} d),e),f) ) we find high values of the
correlation coefficient for the \textit{Satellite1/4}, and
\textit{ROMS1/12} cases.
The suppressive effect of stirring might be dominant
only when stirring is intense, as in the south Benguela.
\citet{Gruber2011} stated that the reduction of biomass due 
to eddies may extend beyond the regions of the most intense 
mesoscale activity, including the offshore areas that we do 
not simulate in this work.

In the following we analyse the bio-physical mechanisms behind 
this negative relationship.

\subsection{What causes the variable biological responses within regions
of distinct dynamical properties?}
\label{subsec:spatial_analysis}

In the following, our analysis is focused on the setting using
\textit{ROMS1/12} as the previous results revealed that the
negative correlation is more robust. Similar results
and conclusions can be obtained from the simulations using the
two other velocity fields (not shown), attesting of the
reliability of our approach (see correlation coefficients and
slopes in Fig. \ref{fig:Phyto_FSLE}).

To understand why simulated chlorophyll-a concentrations differs in
both subsystems, as is the case in satellite observations, we compute
annual budgets of $N, P, Z$ and biological rates (Primary Production $PP$, Grazing and 
Remineralization) in the case of the biological module alone 
(Table \ref{tab.budgets_onlybio}) and when coupled with a 
realistic flow  (Table \ref{tab.budgets_coupled}). Considering the 
biological module alone, we found that $PP$ in the north 
subsystem is slightly higher than in the southern one (4$\%$, 
see also Table \ref{tab.budgets_onlybio} ), essentially due to the differential nutrient 
inputs $\Phi_{N}$. However, when considering the full coupled system 
(hydrodynamic and biology), the latitudinal difference in $PP$
increases significantly (32$\%$, see also Table \ref{tab.budgets_coupled}). 
This latitudinal difference is in agreement with the patterns of $PP$ derived from 
remote-sensed data by \cite{Carr2002}. These results indicate that the flow 
is the main responsible of the difference in PP. Additional 
computations (see \ref{sensitivity_homo}) also confirm the minor effect of the 
biological module ($\Phi_{N}$), as compared with the flow, on the observed 
latitudinal differences in $PP$.

\begin{table}[ht!]
\begin{center}
\begin{tabular}{|l | c | c | c|}
\hline
\hline
\multicolumn{4}{|c|}{Annual budgets only biological system}\\
\hline                                           
\hline                                      
~~~~~~~~~~~~~~~~ &  South    &   North   &  North-South difference ($\%$) \\
\hline  
Nutrients ($mmol N m^{-3}$)  &  821      &   1305    &  37 \\
\hline  
Phytoplankton ($mmol N m^{-3}$) &  57.0    &   57.7     &  1 \\
\hline
Zooplankton ($mmol N m^{-3}$) &  113      &   115    &  2 \\
\hline
Primary Production ($mmol N m^{-3}yr^{-1}$) &  35      &   36    &  4 \\
\hline
Grazing ($mmol N m^{-3}yr^{-1}$)&  33      &   35    &  4 \\
\hline
$\Phi_{N}$ ($mmol N m^{-3}yr^{-1}$) &  28      &   29    &  3 \\
\hline
Remineralization ($mmol N m^{-3}yr^{-1}$)&  7.0      &   7.4    &  4 \\
\hline 
\hline
\end{tabular}
\end{center}
\caption{Budgets of N,P,Z and biological rates (Primary Production, Grazing, $\Phi_{N}$, and remineralization) 
for the biological model.
}
\label{tab.budgets_onlybio}
\end{table}

\begin{table}[ht!]
\begin{center}
\begin{tabular}{|l | c | c | c|}
\hline
\hline
\multicolumn{4}{|c|}{Annual budgets hydrodynamics-biology coupled system}\\
\hline                                           
\hline                                      
~~~~~~~~~~~~~~~~~~~~~~~~~~~ &  South    &   North   &  North-South difference ($\%$) \\
\hline  
Nutrients ($mmol N m^{-3}$)   &  849      &   1937    &  56 \\
\hline  
Phytoplankton  ($mmol N m^{-3}$) &  147    &   198     &  26 \\
\hline
Zooplankton  ($mmol N m^{-3}$) &  231      &   347    &  33 \\
\hline
Primary Production  ($mmol N m^{-3}yr^{-1}$) &  63      &   98    &  32 \\
\hline
Grazing  ($mmol N m^{-3}yr^{-1}$)&  56      &   87    &  35 \\
\hline
$\Phi_{N}$  ($mmol N m^{-3}yr^{-1}$) &  81      &   91    &  10 \\
\hline
Remineralization  ($mmol N m^{-3}yr^{-1}$) &  11      &   18    &  4 \\
\hline 
\hline
\end{tabular}
\end{center}
\caption{Budgets of N,P,Z and biological rates (Primary Production, Grazing, $\Phi_{N}$, and remineralization) 
for the bio-flow coupled model.
}
\label{tab.budgets_coupled}
\end{table}

\begin{figure}[htb]
\begin{center}
\includegraphics[width=0.85\textwidth, height=0.80\textheight, angle=270]{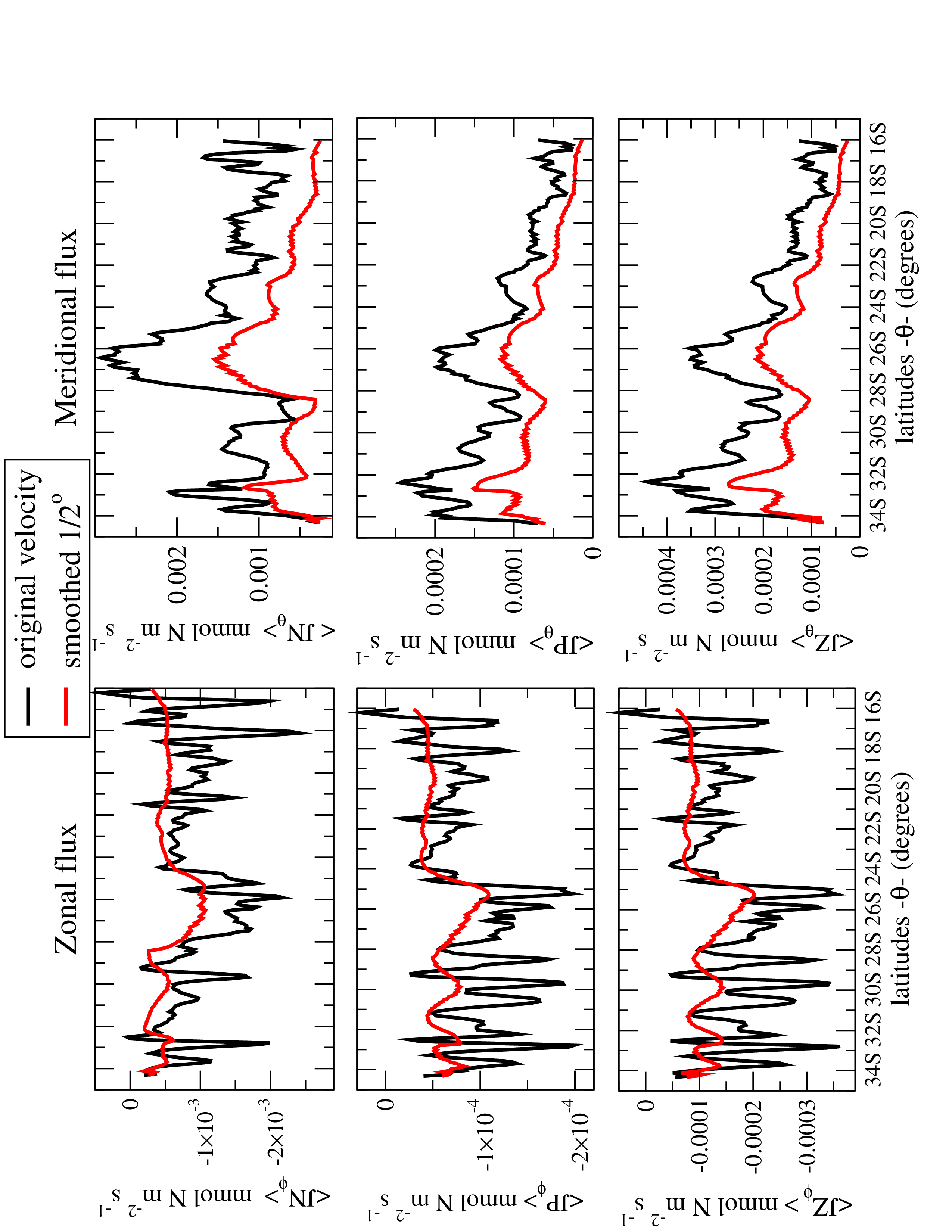}\\
\end{center}
\caption{Zonal mean of zonal and meridional fluxes of $N, P, Z$ concentrations for
the \textit{ROMS1/12} case, averaged from the coast to 3$\degree$ offshore.
 }
\label{fig:flux_roms1_12}
\end{figure}

\citet{Gruber2011}) suggested that the offshore advection 
of plankton biomass enhanced by mesoscale structures might 
be responsible for the suppressive effect of stirring in 
upwelling areas. To test this mechanism, we next analyze 
the net horizontal transport of biological tracers by 
the flow. In  particular, we have computed the zonal, $JC_{\phi}$,
and meridional, $JC_{\theta}$, advective fluxes of $N,P,Z$
(the diffusive fluxes being much smaller):

\begin{eqnarray}
JC_{\phi}(\textbf{x},t)&=&u(\textbf{x},t) C(\textbf{x},t),\label{flux_x}\\
JC_{\theta}(\textbf{x},t)&=&v(\textbf{x},t) C(\textbf{x},t),
\label{flux_y}
\end{eqnarray}
where $u$ and $v$ are the zonal and meridional components of
the velocity field respectively, and with $C$ we denote the N,
P and Z concentrations, all of them given at a specific point
in the 2D-space and time $(\textbf{x},t)$. $JC$ is the flux of
the concentration, $C$, i.e., $JN_{\phi}$ is the zonal flux of
nutrients (eastward positive), $JP_{\theta}$ is the
meridional flux (northward positive) of phytoplankton, and so
on. Annual averages of daily fluxes were computed, and then a
zonal average as a function of the latitude was calculated for
the different coastal bands considered all along this
paper. Fig. \ref{fig:flux_roms1_12} shows these calculations
for the velocity field from \textit{ROMS1/12}, while similar 
results were found for the other data sets (not shown). 
Similar behavior is observed for the fluxes of $N$, $P$ 
and $Z$: zonal fluxes are
almost always negative, so that westward transport dominates,
and meridional fluxes are predominantly positive so that they
are directed to the north. Comparing North and South in the
3$\degree$ coastal band, it is observed that at high latitudes
the zonal flux has larger negative values than at low latitudes,
and the meridional flux presents larger positive values at
higher latitudes. In other words, the northwestward transport
of biological material is more intense in the southern than in
the northern regions, suggesting a higher 'flushing rate'. 
It also suggests that unused nutrients from the southern 
Benguela might be advected toward the northern areas, possibly 
promoting even further the local ecosystem.

To estimate the transport of recently upwelled nutrients 
by LCSs and other mesoscale structures, apart from the mean flow,  
we compute the zonal and meridional fluxes of biological tracers 
using the smoothed \textit{ROMS1/12} velocity field at the spatial 
resolution equivalent to 1/2$\degree$ (see \ref{ape:smooth} for more 
details). The results, plotted in Fig. \ref{fig:flux_roms1_12} (red lines), 
show that in general the fluxes are less intense in the coarser than 
in the finer velocity, indicating that there is a contribution to net transport
due to the submeso- and meso-scale activity.
To estimate the quantitative contribution of mesoscale processes,  we compute
the difference of the fluxes of the different biological tracers $C$ = $N,P,Z$, $Q_{JC}$, in 
the coarser velocity field with respect to the original velocity field. The values 
of $Q_{JC}$ range from 30 to 50$\%$, indicating that the contribution of the mesocale 
to the net transport of the biological concentrations is important. Moreover, the values 
of $Q_{JC}$ are larger in the south than in the north confirming that the mesoscale-induced 
transport is more intense in the south.


\cite{Lachkar2011} showed that mesoscale
processes reduce the efficiency of nutrients utilization by 
phytoplankton due to their influence on residence times. The
longer residence times (i.e. the less mesoscale activity) seem to favor
the accumulation of biomass. To test this effect in our simulations, 
we compute the residence times (RT), defined as the the time interval that a 
particle remains in the coastal trip of 5$\degree$ wide. 
The spatial distribution (not shown) of the annual average of RT indicates 
that the longest RT are found in the north region. In fact, zonal analysis 
reveals that RT has a tendency to increase as the latitude decreases, 
with a mean value in the North equals to 249$days$, and 146$days$ in 
the South. This suggests that regions with weak fluxes are associated with 
long residence times and high growth rate of phytoplankton. On the other 
hand, high mesoscale activity is favoring the northwestward advection which
decreases the residence times, associated to lower growth rate
of plankton.

\begin{figure}[htb]
\begin{center}
\includegraphics[width=0.80\textwidth, height=0.80\textheight]{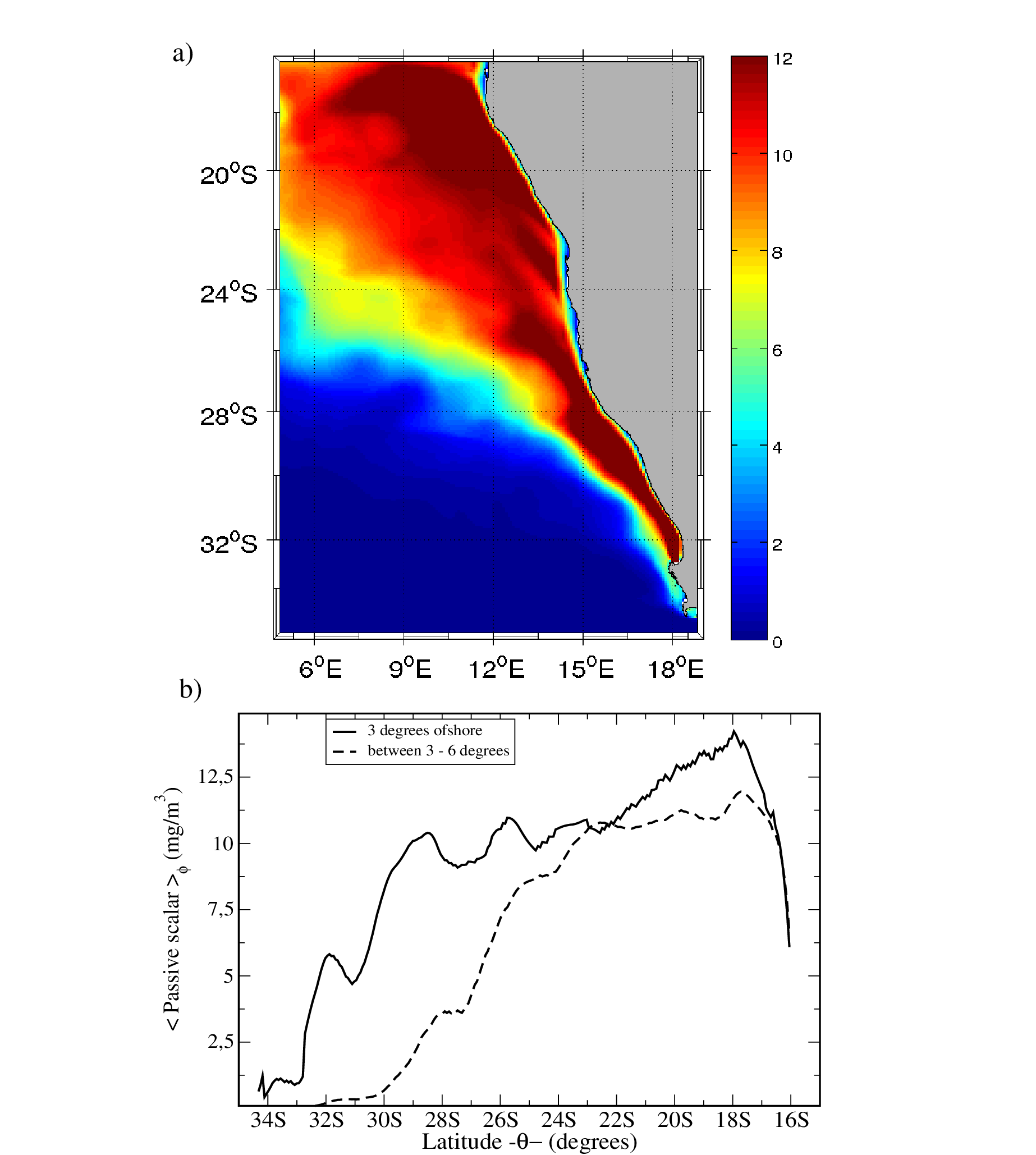}
\end{center}
\caption{a) Spatial distribution of time average of the passive scalar concentration
(see details at the end of subsection \ref{subsec:spatial_analysis}).
b) Comparison of latitudinal profile of time averages of the passive scalar,
as a function of latitude, for zonal average over different coastal bands.}
\label{fig.advecc_sensit}
\end{figure}

This effect and the role of horizontal advection is confirmed 
by performing numerical simulations where no biological dynamics is 
considered. This amounts to solving Eq. (\ref{Eq.biolo1}) with $P=Z=0$ considering 
solely lateral transport, so that $N$ is a passive scalar with  
sources.  In Fig. \ref{fig.advecc_sensit} we see
the results (for the $ROMS1/12$ case, similar for the
other datasets). There is a very small tracer concentration in the
southern domain, and the differences north-south are more
pronounced than the case including the plankton dynamics (see
Fig. \ref{fig:phyto_lat}). This supports further the fact that the main actor 
on the spatial distribution of biomasses is the horizontal transport.

\section{Conclusions}
\label{sec:summary}

We have studied the biological dynamics in the Benguela area
by considering a simple
biological NPZ model coupled with different velocity fields
(satellite and model). Although in a simple framework, a
reduction of phytoplankton concentrations in the coastal
upwelling for increasing mesoscale activity has been
successfully simulated. Horizontal stirring was estimated by
computing the FSLEs and was correlated negatively with
chlorophyll stocks. Similar correlations are found, though not
presented in this manuscript, for the primary production. 
Some recent observational
and modelling studies proposed the ``nutrient leakage" as a
mechanism to explain this negative correlation. Here we argue
that Lagrangian Coherent Structures, mainly mesoscale eddies
and filaments, transport a significant fraction (30-50\%) 
of the recently upwelled nutrients nearshore toward the 
open ocean before being efficiently used by the pelagic food web. 
The fluxes of nutrients and organic matter, due to the mean flow 
and its mesoscale structures, reflect that transport is predominantly 
westward and northward. Biomass is transported towards open 
ocean or to the northern area. In addition to the direct 
effect of transport, primary production is also negatively affected by 
high levels of turbulence, especially in the south Benguela.
Although some studies dealt with 3D effects, we have shown that 2D advection processes
seems to play an important role in this suppressive effect. Our
analysis suggests that the inhibiting effect of the mesoscale
activity on the plankton occurs when the stirring reaches high
levels, as in the south Benguela. However, this effect is not
dominant under certain levels of turbulence. It might indicate that planktonic
ecosystems in oceanic regions with vigorous mesoscale dynamics
can be, as a first approximation, easily modeled just by
including a realistic flow field. The small residence times of waters 
in the productive area will smooth out all the other neglected biological 
factors in interaction.

Our findings confirm the unexpected role that mesoscale activity has on
biogeochemical dynamics in the productive coastal upwelling. Strong vertical
velocities are known to be associated with these physical structures and they
might have another direct effect by transporting downward rich nutrient waters
below the euphotic zone. Further studies are needed such as 3D realistic
modelling that take into account the strong vertical dynamics in upwelling
regions to test the complete mechanisms involved.

\section*{Acknowledgments}

I.H-C was supported by a FPI grant from MINECO to visit LEGOS. We acknowledge support 
from MINECO and FEDER through projects FISICOS (FIS2007-60327) and 
ESCOLA (CTM2012-39025-C02-01). V. G.
thanks CNES funding through Hiresubcolor project. We are also grateful to J. Sudre
for providing us velocity data sets both from ROMS and from the combined satellite
product. Ocean color data were produced by the SeaWiFS project at GES and were 
obtained from DAAC.

\appendix
\section{Sensitivity analysis}

A number of numerical experiments were done to investigate the sensitivity of the coupled
bio-physical model with respect to different variables.

\subsection{Sensitivity with respect to different spatial resolution of the velocity field}
\label{ape:smooth} In this experiment we used a velocity field
from ROMS1/12 smoothed out towards a resolution 1/4$\degree$,
and to be compared with $ROMS1/4$ and $ROMS1/12$ at their
original spatial resolution. We coarse-grained the velocity
field with a convolution kernel weighted with a local
normalization factor, and keeping the original resolution for
the data so that land points are equally well described as in
the original data. The coarsening kernel with scale factor $s$,
$k_{s}$, is defined as:

\begin{equation}
k_{s}(x,y)=e^{-\frac{(x^{2}+y^{2})}{2s^{2}}}.
\label{eq.kernel}
\end{equation}

To avoid spurious energy dump at land points we have introduced
a local normalization weight given by the convolution:
$k_{s}(x,y)*M(x,y)$, where $M(x,y)$ is the sea mask. For points
far from the land the weight is just the normalization of
$k_{s}$, and for points surrounded by land the weight takes the
contribution from sea points only. Thus $v_{s}$, the velocity
field coarsened by a scale factor $s$, is obtained from the
original velocity field $v$ as:

\begin{equation}
v_{s}=\frac{k_{s}*v}{k_{s}*M}.
\label{eq.conv}
\end{equation}

\begin{figure}[htb]
\begin{center}
\includegraphics[width=0.80\textwidth, height=0.80\textheight]{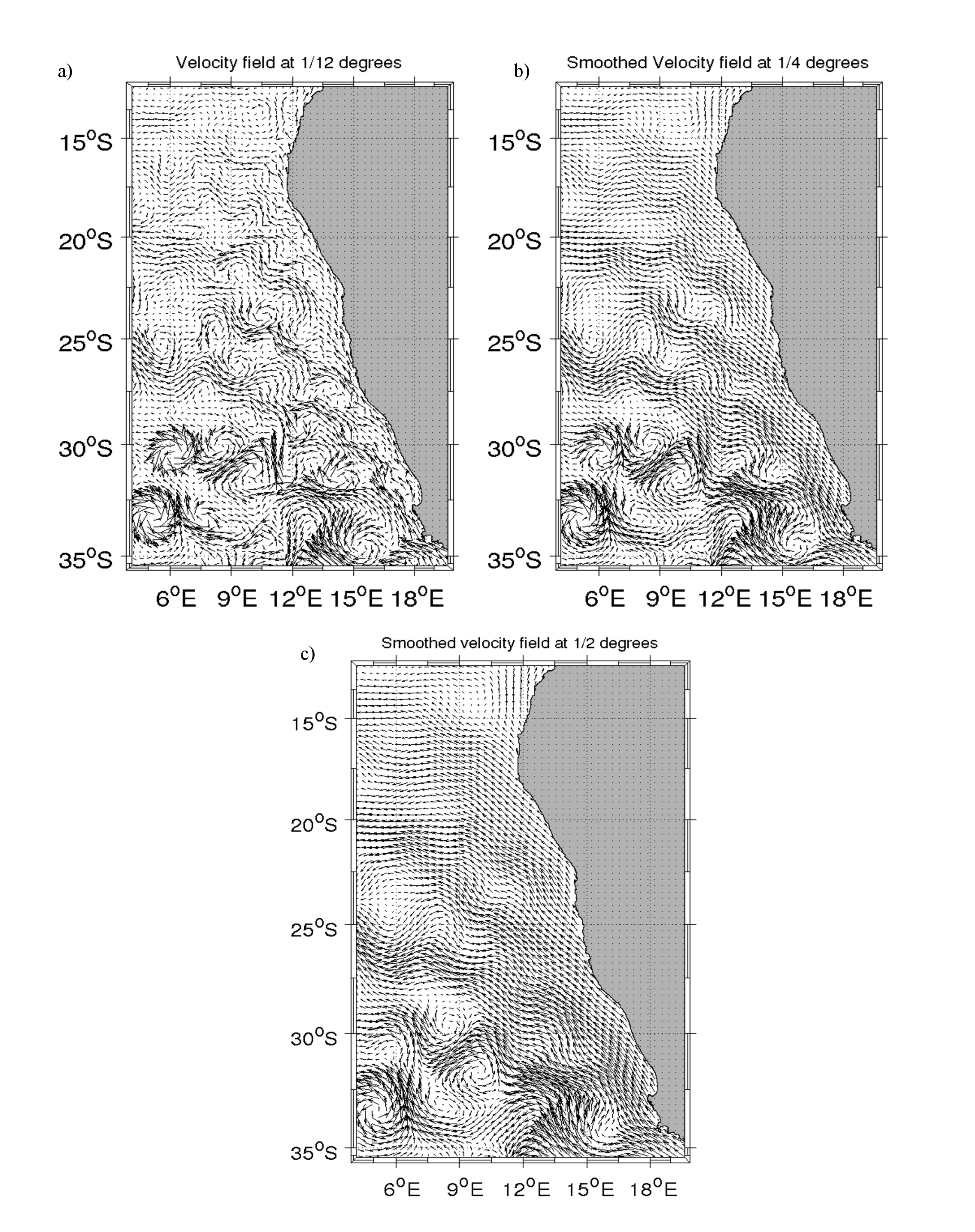}
\end{center}
\caption {Vectors of a velocity field from $ROMS1/12$: a) at original resolution. b) smoothed by
a scale factor of $s$=3, obtaining and equivalent spatial resolution of 1/4$\degree$,
c) smoothed by a scale factor of s=6, obtaining and equivalent spatial resolution of
1/2$\degree$. The snapshots correspond to day 437 of the simulation.
}
\label{fig.smooth_vel}
\end{figure}

\begin{figure}[htb]
\begin{center}
\includegraphics[width=0.99\textwidth, height=0.65\textheight]{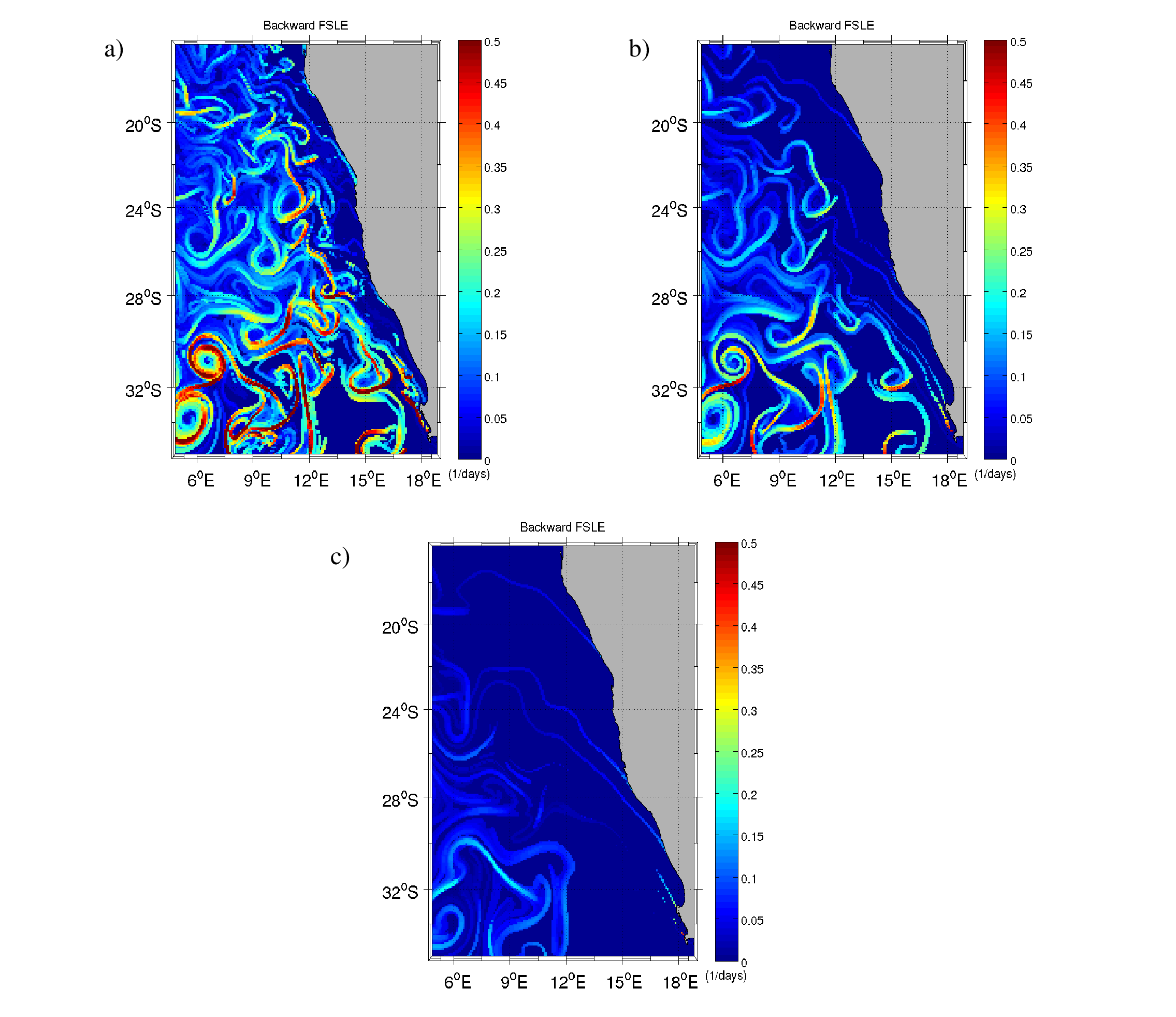}
\end{center}
\caption {Snapshots of spatial distributions of FSLEs backward 437 days in time
starting from day 437 of $ROMS1/12$ at the same FSLE grid resolution of 1/12$\degree$, and
using the velocity fields at different resolutions: a) at original resolution 1/12$\degree$.
b) smoothed velocity field at equivalent 1/4$\degree$ and c) smoothed velocity field
at equivalent 1/2$\degree$.}
\label{fig.FSLEsmooth_vel}
\end{figure}

In Fig. \ref{fig.smooth_vel} we compare two \textit{ROMS1/12}
smoothed velocity fields at scales $s$=3 and $s$=6 (with an
equivalent spatial resolution 1/4$\degree$ and 1/2$\degree$,
respectively) with the original velocity field from
\textit{ROMS1/12}. It is clear that the circulation pattern is
smoothed as $s$ is increased. The FSLE computations using these
smoothed velocity fields are shown in Fig
\ref{fig.FSLEsmooth_vel}. When the spatial resolution is
reduced to $1/4\degree$ the FSLEs and small-scale contributions
decrease, but the main global features remain, as indicated in
the study by \cite{HernandezCarrasco2011}. Further coarsening
to $1/2\degree$ smoothes most of the structures except the most
intense ones.

\begin{figure}[htb]
\begin{center}
\includegraphics[width=0.65\textwidth, height=0.40\textheight]{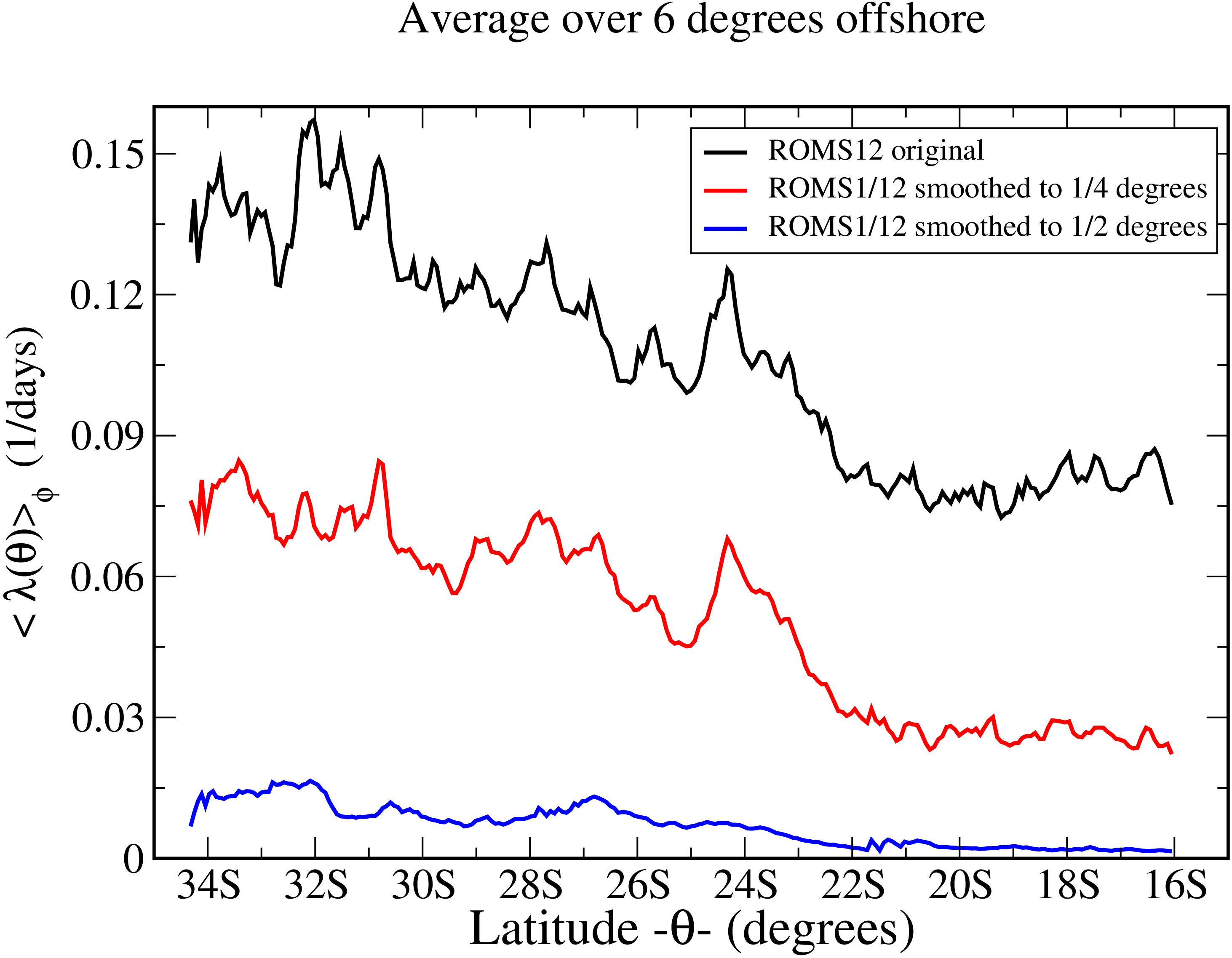}
\end{center}
\caption {Latitudinal profile of the zonal mean values of annual averaged
backward FSLEs (51 snapshots weekly separated)
at the same FSLE grid resolution of 1/12$\degree$, and using different smoothed velocity fields.
}
\label{fig.comparison_FSLEsmooth}
\end{figure}

The latitudinal variations of the zonal averages performed on
the time averages of the FSLE maps plotted in Fig.
\ref{fig.FSLEsmooth_vel} are compared in
Fig.\ref{fig.comparison_FSLEsmooth}. The mean FSLEs values
strongly diminish when the velocity resolution is sufficiently
smoothed out. This is due to the progressive elimination of
mesoscale structures that are the main contributors to stirring
processes. Also the latitudinal variability of stirring
diminishes for the very smoothed velocity field (blue line in
Fig. \ref{fig.comparison_FSLEsmooth} ). Thus, latitudinal
differences of stirring in the Benguela system are likely to be
related to mesoscale structures (eddies, filaments, fronts,
etc.) contained in the velocity fields.

\begin{figure}[htb]
\begin{center}
\includegraphics[width=0.99\textwidth, height=0.65\textheight]{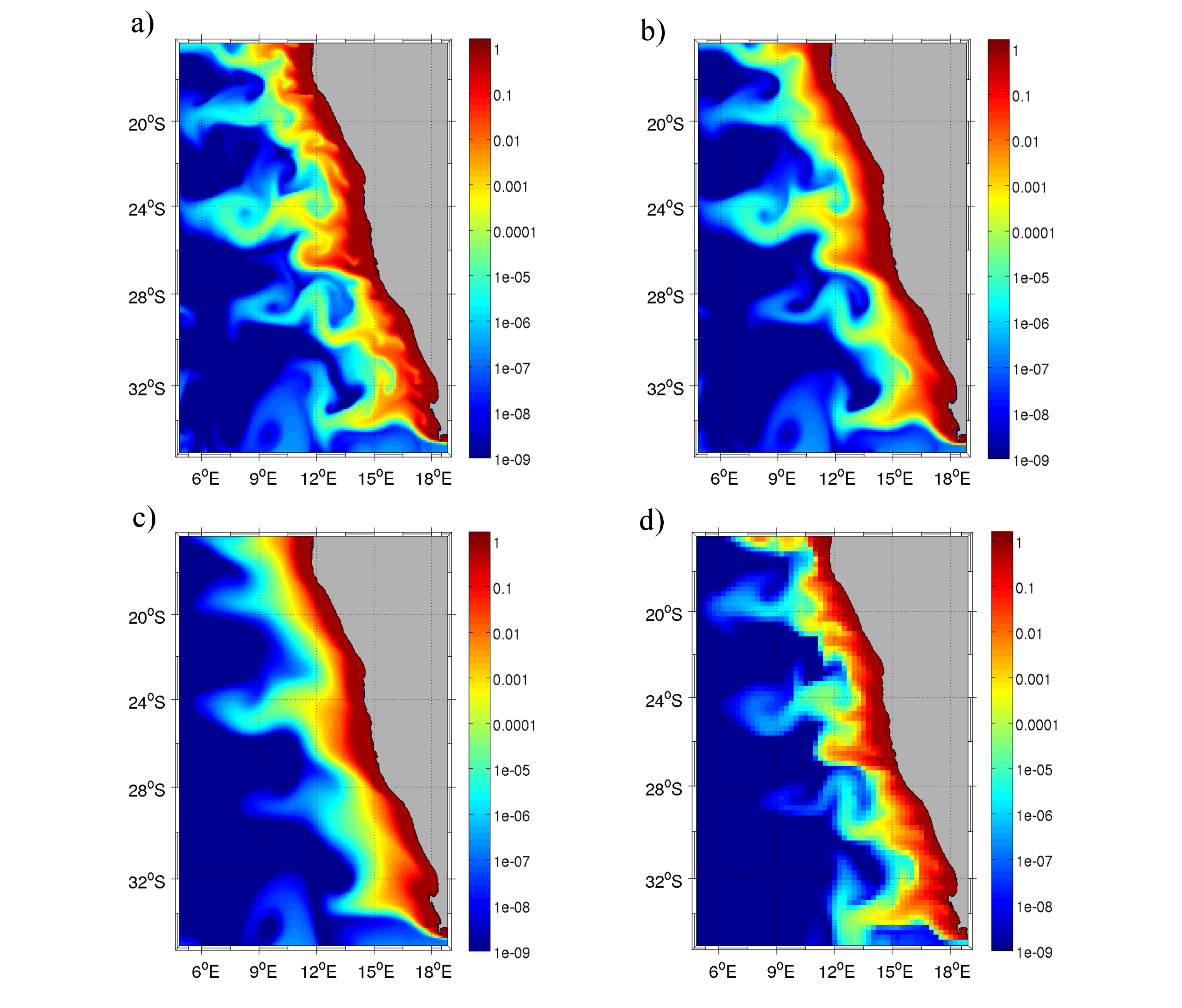}
\end{center}
\caption {Snapshots of simulated chlorophyll field using different velocity fields:
a) $ROMS1/12$ at original resolution 1/12$\degree$, b) smoothed $ROMS1/12$
velocity field at equivalent 1/4$\degree$, c) smoothed $ROMS1/12$ velocity field at equivalent 1/2$\degree$,
and d) $ROMS1/4$ at original resolution 1/4$\degree$. The units of the colorbar are $mg/m^{3}$.
}
\label{fig.Phyto_smooth_vel}
\end{figure}

We have also computed the phytoplankton using these smoothed velocity fields.
Some instantaneous spatial distributions can be seen in Fig \ref{fig.Phyto_smooth_vel}.
The filaments of phytoplankton disappear in the very smoothed velocity field (1/2$\degree$).
The spatial distribution of the annual average of phytoplankton concentrations for
the different velocity field shows, however, quite similar patterns (not shown).

In the time series of $N$, $P$ and $Z$ budgets for the coarser velocities
 one observes the same behavior as for the original velocity field (not shown).

\subsection{Sensitivity with respect to different parameterization of the coastal upwelling of nutrients.}
\label{sensitivity_homo}

\begin{figure}[htb]
\begin{center}
\includegraphics[width=0.55\textwidth, height=0.52\textheight, angle=270]{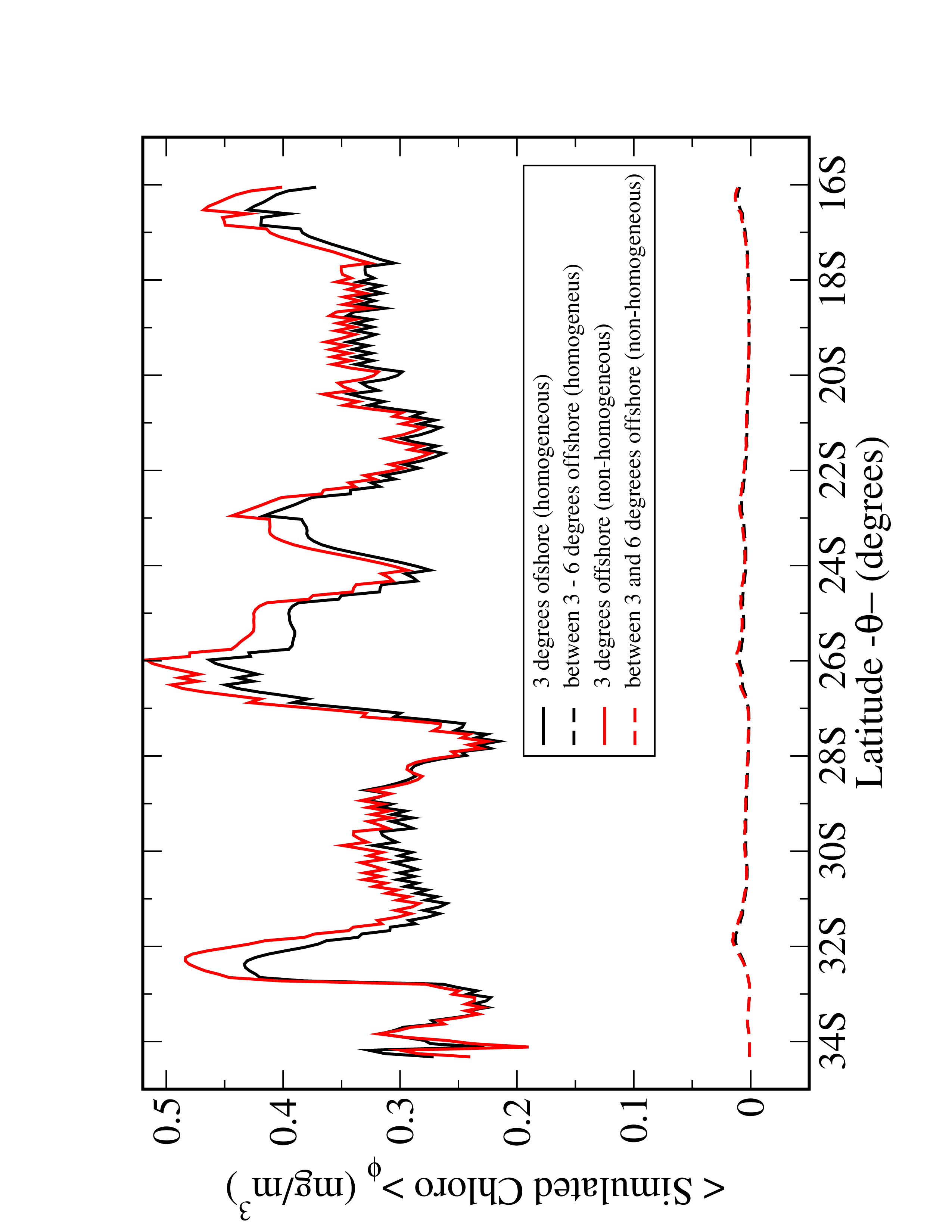}
\end{center}
\caption{Comparison between zonal average on different coastal bands of annual time average
of simulated chlorophyll, using homogeneous upwelling and the non-homogeneous 
upwelling cells described in Fig. \ref{fig:cells}.
}
\label{fig:npz_homocells}
\end{figure}

In section \ref{bio} we mimicked coastal upwelling of nutrient
via a source term in the nutrients equation which is determined
by the function $S$, and was considered spatiotemporally
variable. Here we explore the plankton dynamics using spatially
and temporally homogeneous upwelling along the coast. $S$ is
fixed to an average value $S= 0.1 \ day^{-1}$ along the coast
at any time. In Fig. \ref{fig:npz_homocells} we show the annual
average of $P$ for the $ROMS1/12$ (top panel), and the
comparisons with the inhomogeneous case for the zonal mean
(bottom panel). Therefore, this test suggests that the way we
simulate vertical mixing along the coast has not a large effect
on the 2D biological dynamics, which will be mainly determined
by the interplay with horizontal advection.

%

\end{document}